\documentclass{aa}

\usepackage{graphicx}
\usepackage{natbib}

\newcommand{\id}{\mbox{$\mathrm{d^{-1}}$}}

\newcommand{\kms}{\mbox{$\mathrm{km\,s^{-1}}$}}

\newcommand{\T}{\mbox{$T_\mathrm{0}$}}

\newcommand{\Porb}{\mbox{$P_{\rm orb}$}}
\newcommand{\Line}[3]{\Ion{#1}{#2}\,\,$\lambda$#3}

\newcommand{\Ion}[2]{#1{\,\scriptsize #2}}
\newcommand{\Ha}{\mbox{${\mathrm H\alpha}$}}
\newcommand{\Hb}{\mbox{${\mathrm H\beta}$}}
\newcommand{\Hg}{\mbox{${\mathrm H\gamma}$}}

\begin{document}

\title{HS\,0139+0559, HS\,0229+8016, HS\,0506+7725 and HS\,0642+5049:
\\ Four new long-period cataclysmic variables\thanks{Based on
observations obtained at the German-Spanish Astronomical Center, Calar
Alto, operated by the Max-Planck-Institut f\"{u}r Astronomie,
Heidelberg, jointly with the Spanish National Commission for
Astronomy,
on observations made at the 1.2m telescope, located at Kryoneri
Korinthias, and owned by the National Observatory of Athens, Greece, 
on observations made with the  OGS telescope, operated on
the island of Tenerife by the European Space Agency, in the
Spanish Observatorio del Teide of the IAC.
}}

\author{A. Aungwerojwit\inst{1}\and 
        B.T. G\"ansicke\inst{1}\and
        P. Rodr\'iguez-Gil\inst{1,2}\and
        H.-J. Hagen\inst{3}\and
        E.T. Harlaftis\inst{4}\and
        C. Papadimitriou\inst{{5},{6}}\and
        H. Lehto\inst{{7},{8}}\and
        S. Araujo-Betancor\inst{9}\and
        U. Heber\inst{10}\and
        R.E. Fried\inst{11}\and
        D. Engels\inst{3}\and
	S. Katajainen\inst{8}
        }
\authorrunning{Aungwerojwit et al.}
\titlerunning{Four new long-period cataclysmic variables}

\offprints{A. Aungwerojwit, \\ e-mail: A.Aungwerojwit@warwick.ac.uk}

\institute{
   Department of Physics, University of Warwick, Coventry CV4 7AL, UK
\and
   Instituto de Astrof\'isica de Canarias, 38200 La Laguna, Tenerife, Spain
\and 
   Hamburger Sternwarte, Universit\"at Hamburg, Gojenbergsweg
   112, 21029 Hamburg, Germany
\and
   \mbox{Institute of Space Applications and Remote Sensing,
   National Observatory of Athens, P.O. Box 20048, Athens 11810, Greece}
\and
   Institute Astronomy and Astrophysics, 
   National Observatory of Athens, P.O. Box 20048, Athens 11810, Greece
\and
   Department of Astrophysics, Astronomy and Mechanics, University of Athens, 
   157 84 Zografos, Athens, Greece
\and
   Department of Physics, FIN-20014 University of Turku, Finland
\and
   Tuorla Observatory, University of Turku, FIN-21500 Piikki\"o, Finland
\and
   Space Telescope Science Institute, 3700 San Martin Drive, Baltimore, 
   MD21218, USA
\and
   Astronomisches Institut der Universit\"at Erlangen,
   Germany
\and
   Braeside Observatory, PO Box906, Flagstaff, AZ 86002, USA
   }

\date{Received \underline{\hskip2cm} ; accepted }

\abstract {We present time-resolved optical spectroscopy and
  photometry of four relatively bright ($V\sim14.0-15.5$) long-period
  cataclysmic variables (CVs) discovered in the Hamburg Quasar Survey:
  HS\,0139+0559, HS\,0229+8016, HS\,0506+7725 and HS\,0642+5049. Their
  respective orbital periods, $243.69\pm0.49$ min, $232.550\pm0.049$
  min, $212.7\pm0.2$ min and $225.90\pm0.23$ min are determined from
  radial velocity and photometric variability studies. HS\,0506+7725
  is characterised by strong Balmer and He emission lines,
  short-period ($\sim10-20$\,min) flickering and weak X-ray emission
  in the ROSAT All Sky Survey. The detection of a deep low state
  ($B\simeq18.5$) identifies HS\,0506+7725 as a member of the VY\,Scl
  stars.  HS\,0139+0559, HS\,0229+8016 and HS\,0642+5049 display
  thick-disc like spectra and no or only weak flickering activity.
  HS\,0139+0559 and HS\,0229+8016 exhibit clean quasi-sinusoidal
  radial velocity varations of their emission lines but no or very
  little orbital photometric variability. In contrast, we detect no
  radial velocity variation in HS\,0642+5049 but a noticeable orbital
  brightness variation. We identify all three systems either as
  UX\,UMa-type novalike variables or as Z\,Cam-type dwarf novae. Our
  identification of these four new systems underlines that the
  currently known sample of CVs is rather incomplete
  even for bright objects. The four new systems add to the clustering of
  orbital periods in the 3--4\,h range found in the sample of HQS
  selected CVs, and we discuss the large incidence of magnetic CVs and
  VY\,Scl/SW\,Sex stars found in this period range among the known
  population of CVs.
\keywords{stars: binaries: close -- stars: individual: HS\,0139+0559,
  HS\,0229+8016, HS\,0506+7725, HS\,0642+5049 -- stars: novae,
  cataclysmic variables} }

\maketitle

\begin{figure*}
\includegraphics[width=4.4cm]{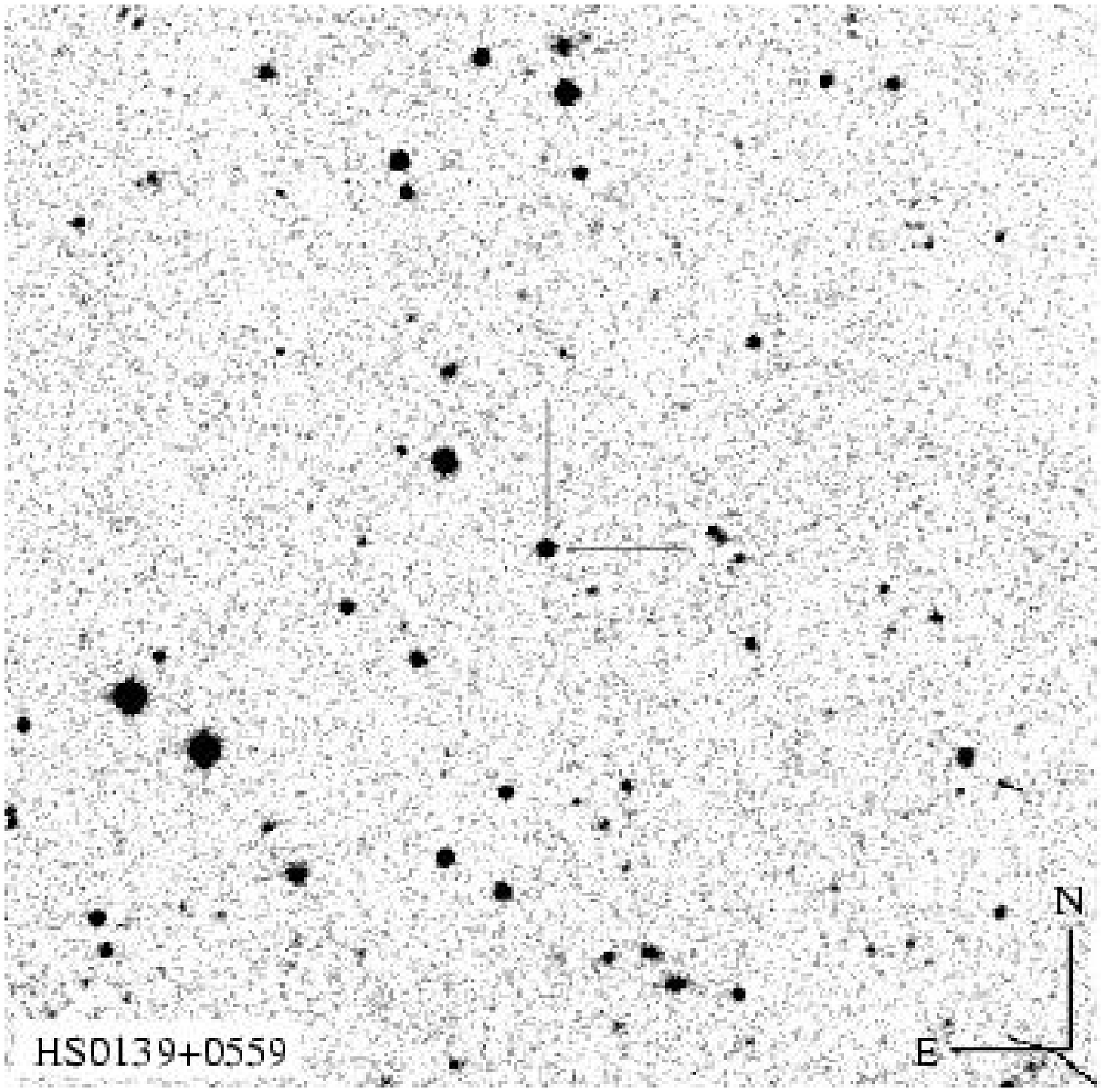}
\hfill
\includegraphics[width=4.4cm]{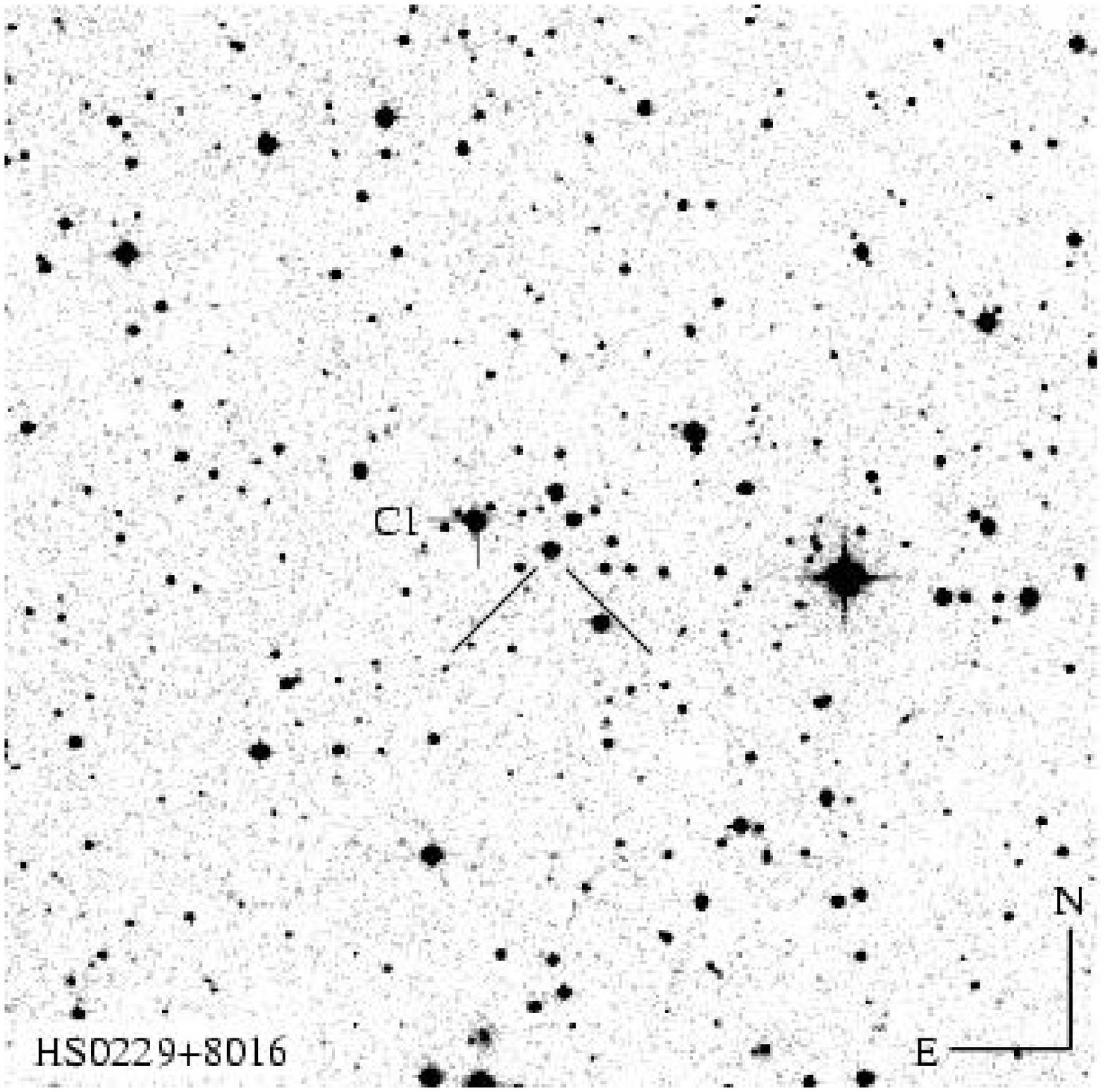}
\hfill
\includegraphics[width=4.4cm]{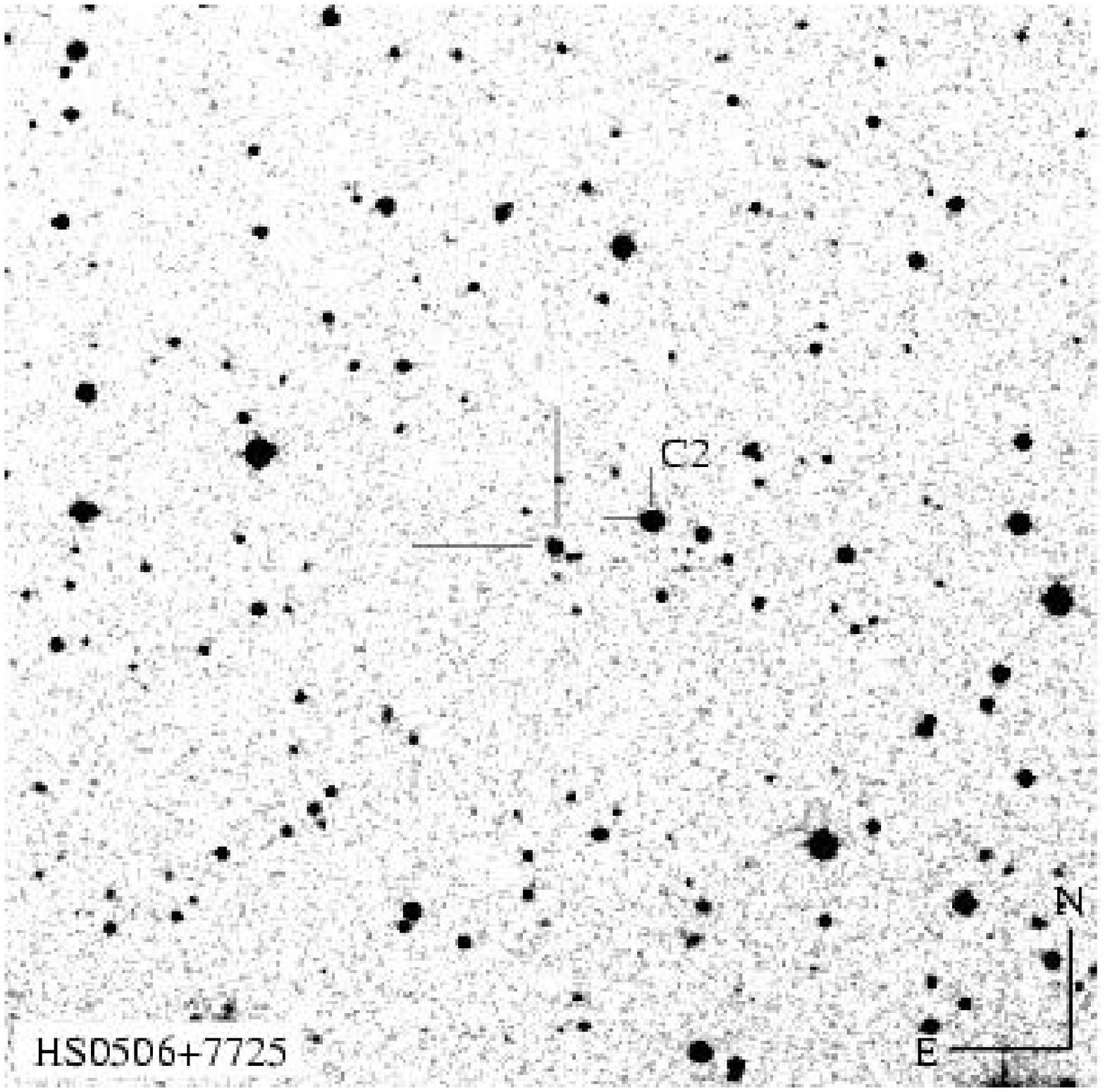}
\hfill
\includegraphics[width=4.4cm]{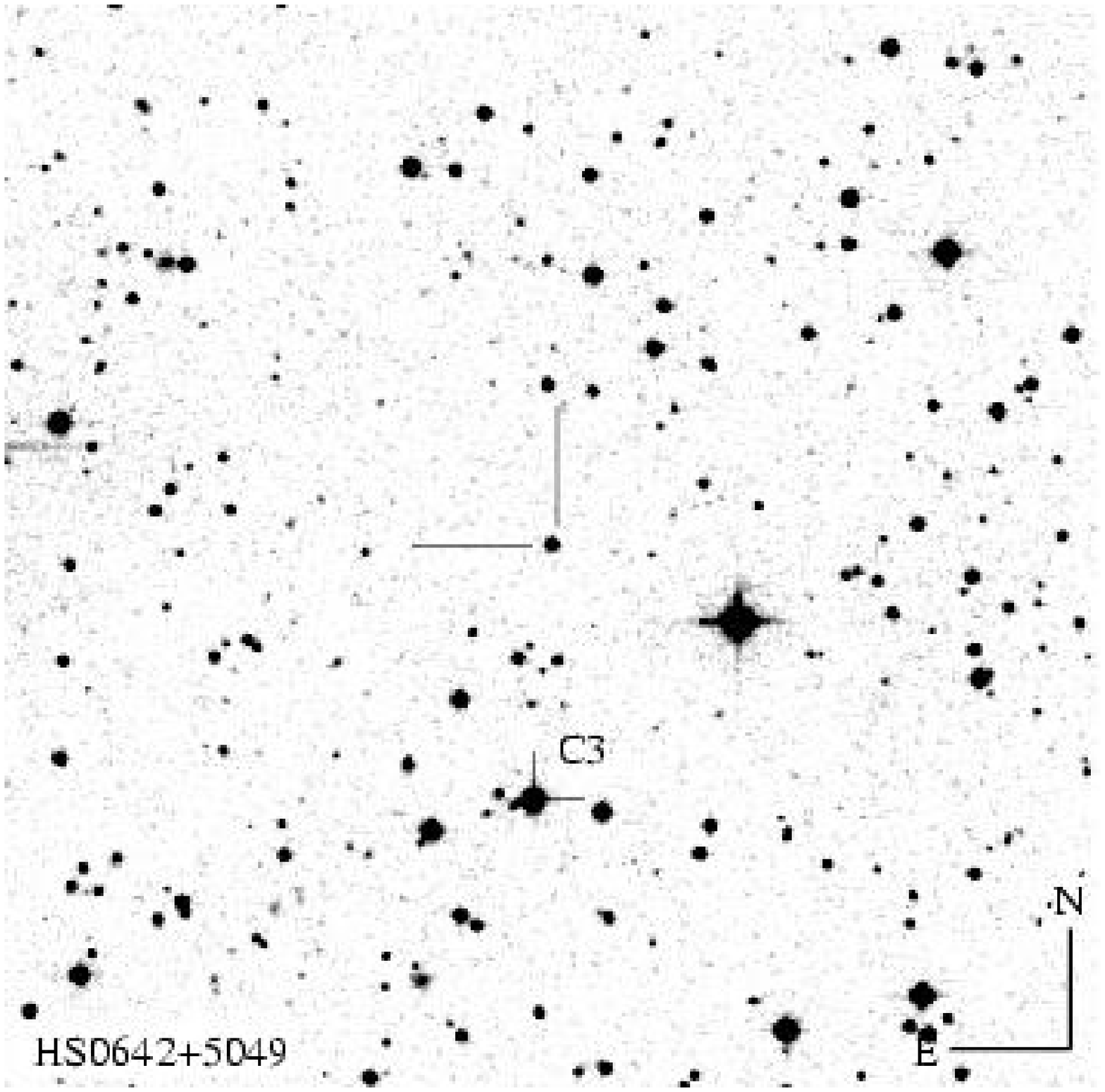}
\caption{\label{f-fc} $10\arcmin\times10\arcmin$ finding charts for
HS\,0139+0559, HS\,0229+8016, HS\,0506+7725 and HS\,0642+5049 obtained
from the Digitized Sky Survey. The J2000 coordinates of the CVs are
($\alpha=01^\mathrm{h}41^\mathrm{m}40.0^\mathrm{s}$,
$\delta=+06\degr14\arcmin36.8\arcsec$),
($\alpha=02^\mathrm{h}35^\mathrm{m}58.3^\mathrm{s}$,
$\delta=+80\degr29\arcmin44.0\arcsec$),
($\alpha=05^\mathrm{h}13^\mathrm{m}36.4^\mathrm{s}$,
$\delta=+77\degr28\arcmin42.1\arcsec$) and 
($\alpha=06^\mathrm{h}46^\mathrm{m}19.60^\mathrm{s}$,
$\delta=+50\degr45\arcmin48.0\arcsec$), respectively.
}
\end{figure*}

\section{Introduction}
Standard models for the population of cataclysmic variables (CVs)
predict that the vast majority of all CVs should have short orbital periods,
$\Porb<2$\,h \citep[e.g.][]{kolb93-1, howelletal97-1}, and a space
density of $2\times10^{-5}-2\times10^{-4}\,\mathrm{pc^{-3}}$
\citep{ritter+burkert86-1, dekool92-1, politano96-1}. These
predictions are in contrast with the properties of the observed
population of galactic CVs, with an estimated space density of
$\sim6\times10^{-6}\,\mathrm{pc^{-3}}$ \citep{ringwald96-1,
araujo-betancoretal05-2} and an apparent lack of short-period
systems. Possible reasons for these discrepancies are uncertainties in the
theory of CV evolution \citep[e.g.][]{king88-1, schenker+king02-1,
andronovetal03-1, barker+kolb03-1, taametal03-1}, but also
observational selection effects in the known CV population 
\citep[e.g][]{downes86-1, ringwald96-1, gaensicke05-1}.

\section{CVs in the Hamburg Quasar Survey}
About 75\% of all known CVs have been discovered either because of
their variability or because of their X-ray emission, with a strong
dominance of dwarf novae and classical nova in the first group and
magnetic CVs in the second group \citep{gaensicke05-1}. It is
therefore clear that CVs which are characterised by infrequent
outbursts and/or low-amplitude variability, as well as lacking strong
X-ray emission, will be underrepresented in the currently known
CV population. Hypothetically, such objects could make up for the
large number of predicted short-period CVs with low mass transfer
rates. 

The primary purpose for our project identifying new CVs in Hamburg
Quasar Survey (HQS) is to establish a large homogeneously selected
sample of systems that overcomes previous observational biases, and
that can subsequently be used to test our understanding of CV
evolution \citep{gaensickeetal02-2}. The HQS, an objective-prism
survey, has been carried out with the 0.8\,m Schmidt telescope at
Calar Alto Observatory to search for bright quasars in the northern
sky at high galactic lattitudes, $\delta > 0\degr$ and $|b| > 20\degr$,
covering $\approx13\,600\,\mathrm{deg^{2}}$, with a dynamic range of
$13 \la B \la 18.5$ \citep{hagenetal95-1}. The photographic plates
covered a spectral range of $\sim3400-5400$\,\AA\ with a resolution
of $\sim45$\,\AA\ at \Hb.

Our CV candidate selection made use of a property that has never been
systematically exploited before: the \textit{spectroscopic} hallmark
of CVs, i.e. the presence of noticeable emission lines in most
CVs. The CV candidate selection was carried out by visually inspecting
48708 HQS prism spectra for the presence of Balmer emission
lines. In order to test the efficiency of this method, we applied the
same procedure to the subset of 84 previously known CVs
(\citealt{downesetal01-1}, as of July 2001) that are contained in the
HQS spectral data base. We positively recovered $\simeq90\%$ of the
known short-period ($\Porb<2$\,h) CVs, including prominent dwarf novae
 such as SW\,UMa or T\,Leo (the latter one has a
rather long outburst cycle of $\sim420$\,d), as well as magnetic CVs
such as AN\,UMa or ST\,LMi. The fraction of recovered systems drops to
$\simeq40\%$ for long-period systems, with the largest fraction of
missed identifications being novalike variables with weak or no Balmer
emission lines. In total, 62\% of the previously known CVs having an
HQS prism spectrum were positively identified by our selection
method. We concluded from this test that the HQS should be
very efficient in finding CVs below the period gap as long as they have
similar spectroscopic properties to the previously known systems, i.e.
Balmer emission lines with equivalent widths $>10$\AA\ in \Hb. The
decrease in detection efficiency for long-period systems has been
compensated to some extent by follow-up programs investigating hot stars in the
HQS, delivering a number of new CVs with weak emission lines
\citep{heberetal91-1}.

In total, 53 new CVs were identified within this project, and and
substantial observational effort has been invested to determine the
properties of these systems. To date, 42 HQS CVs have had their
orbital period measured. Despite its good sensitivity for short-period
CVs, only a small number of new short-period CVs has been found, and
those that have been found fully confirm the expected properties:
low-amplitude variability and/or long outburst recurrence times, e.g.
KV\,Dra (HS\,1449+6415, \citealt{nogamietal00-1}); HS\,2331+3905
\citep{araujo-betancoretal05-1}, DW\,Cnc (HS\,0756+1624,
\citealt{rodriguez-giletal04-1}), or HS\,2219+1824
\citep{rodriguez-giletal05-1}. The majority of the newly identified
systems have orbital periods above the $2-3$\,h period gap,
including rarely outbursting dwarf novae such as GY\,Cnc
(HS\,0907+1902, \citealt{gaensickeetal00-2}) or RX\,J0944.5+0357
(HS0941+0411, \citealt{mennickentetal02-1}); magnetic CVs with
relatively weak X-ray emission, such as 1RXS\,J062518.2+733433
(HS0618+7336, \citealt{araujo-betancoretal03-2}), RX\,J1554.2+2721
(HS\,1552+2730, \citealt{thorstensen+fenton02-1, gaensickeetal04-3})
and HS\,0943+1404 \citep{rodriguez-giletal05-2}; and half a dozen new
SW\,Sextantis stars \citep{gaensickeetal02-3, szkodyetal01-1, 
rodriguez-giletal04-2, rodriguez-gil05-1}.

While a full discussion of the implications that our search for CVs in
the HQS has for our understanding of the galactic CV population has to
await the characterization of the full HQS CV sample, an important
preliminary statement that will not substantially change is
\textit{there is no substantial population of nearby short-period CVs
that resemble the known template systems}. Phrased differently, if the
large population of short period CVs predicted by theory exists, the
majority of these systems must look different from the known
short-period systems, i.e. have weak emission lines and/or
substantially redder continua. Interestingly, the Sloan Digital Sky
Survey (SDSS) is discovering a number of CVs with a steep Balmer
decrement and in which the white dwarf dominates the optical emission, a
clear sign of low mass transfer rates\citep[][and references
therein]{szkodyetal05-1}. However, most of these systems are very
faint, $g\simeq19-20$, implying that they are distant ($d>100$\,pc)
and not intrinsically numerous anywhere near the numbers predicted by
theory. 

An unexpected finding of our search for CVs in the HQS has been the
identification of a large number of systems with orbital periods in
the range $3-4$\,h. Here, we report the discovery of four additional
CVs in that period range, HS\,0139+0559, HS\,0229+8016, HS\,0506+7725
and HS\,0642+5049. Despite being close in orbital period, these
systems differ dramatically in their observed characteristic, and we
discuss in detail the properties of the known CVs in the $3-4$\,h period range.

\section{Observations and Data Reduction}

\subsection{Spectroscopy}
\begin{figure*}
\begin{minipage}[t]{\columnwidth}
\centerline{\includegraphics[angle=-90,width=\columnwidth]{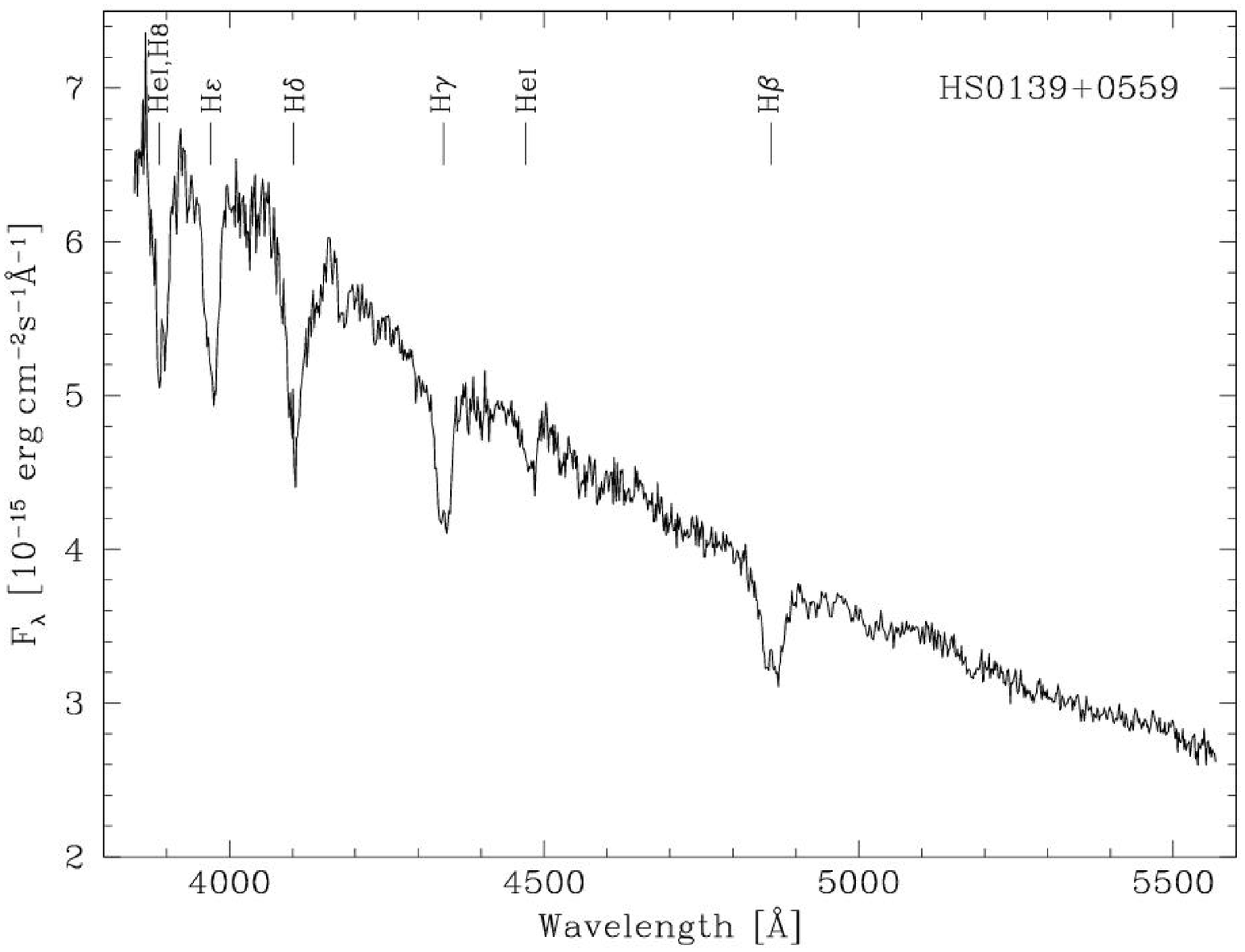}}
\caption[]{\label{f-idspec_hs0139} 
Identification spectrum of HS\,0139+0559 obtained at the Calar Alto
3.5\,m telescope on 1989 January 22.}
\end{minipage}
\hfill
\begin{minipage}[t]{\columnwidth}
\centerline{\includegraphics[angle=-90,width=\columnwidth]{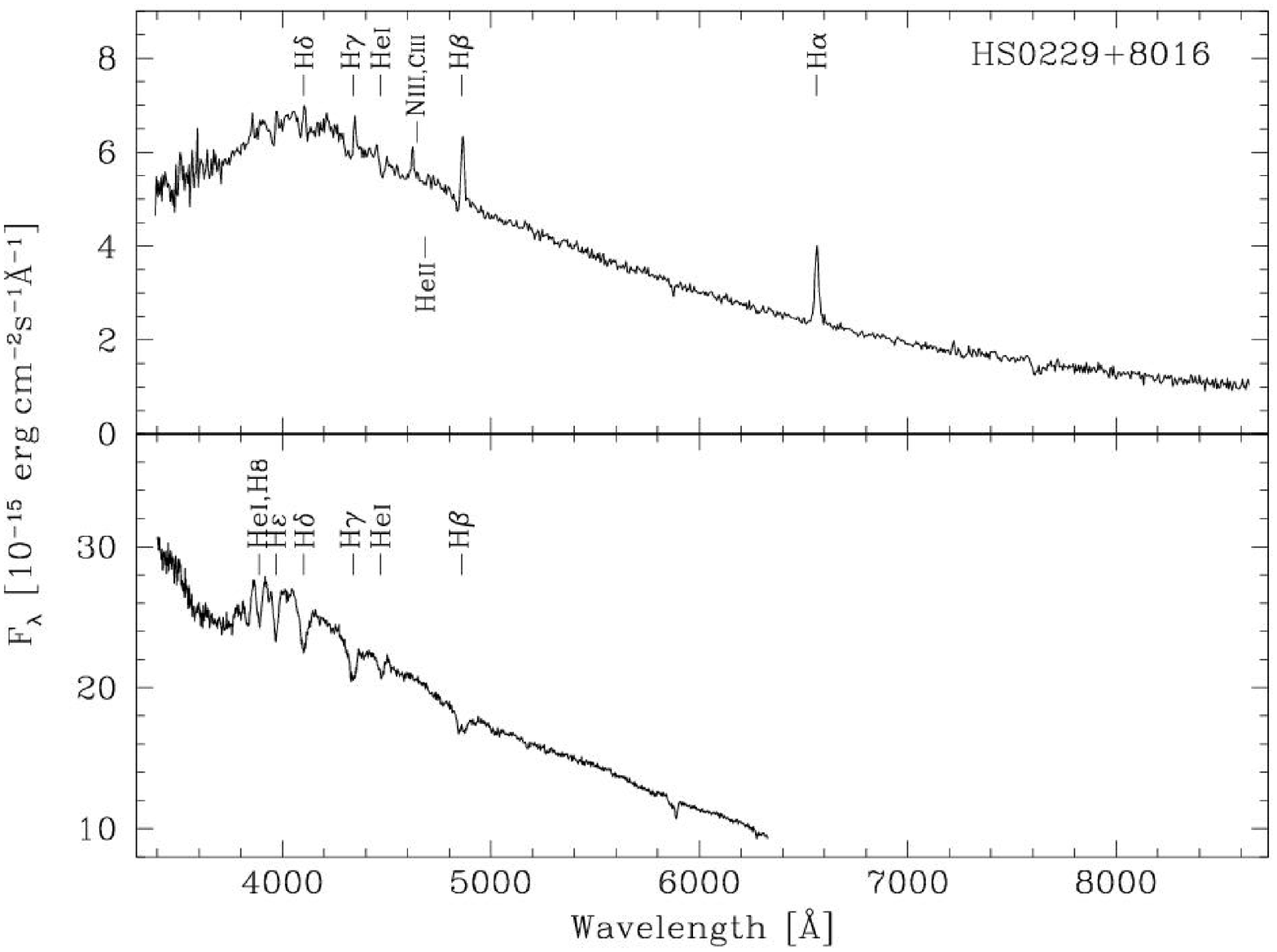}}
\caption[]{\label{f-idspec_hs0229} Identification spectra of
  HS\,0229+8016 obtained at the Calar Alto 2.2\,m telescope on 1992
  August 8 (top panel) and 1998 February 2 (bottom panel).}
\end{minipage}

\smallskip
\begin{minipage}[t]{\columnwidth}
\centerline{\includegraphics[angle=-90,width=\columnwidth]{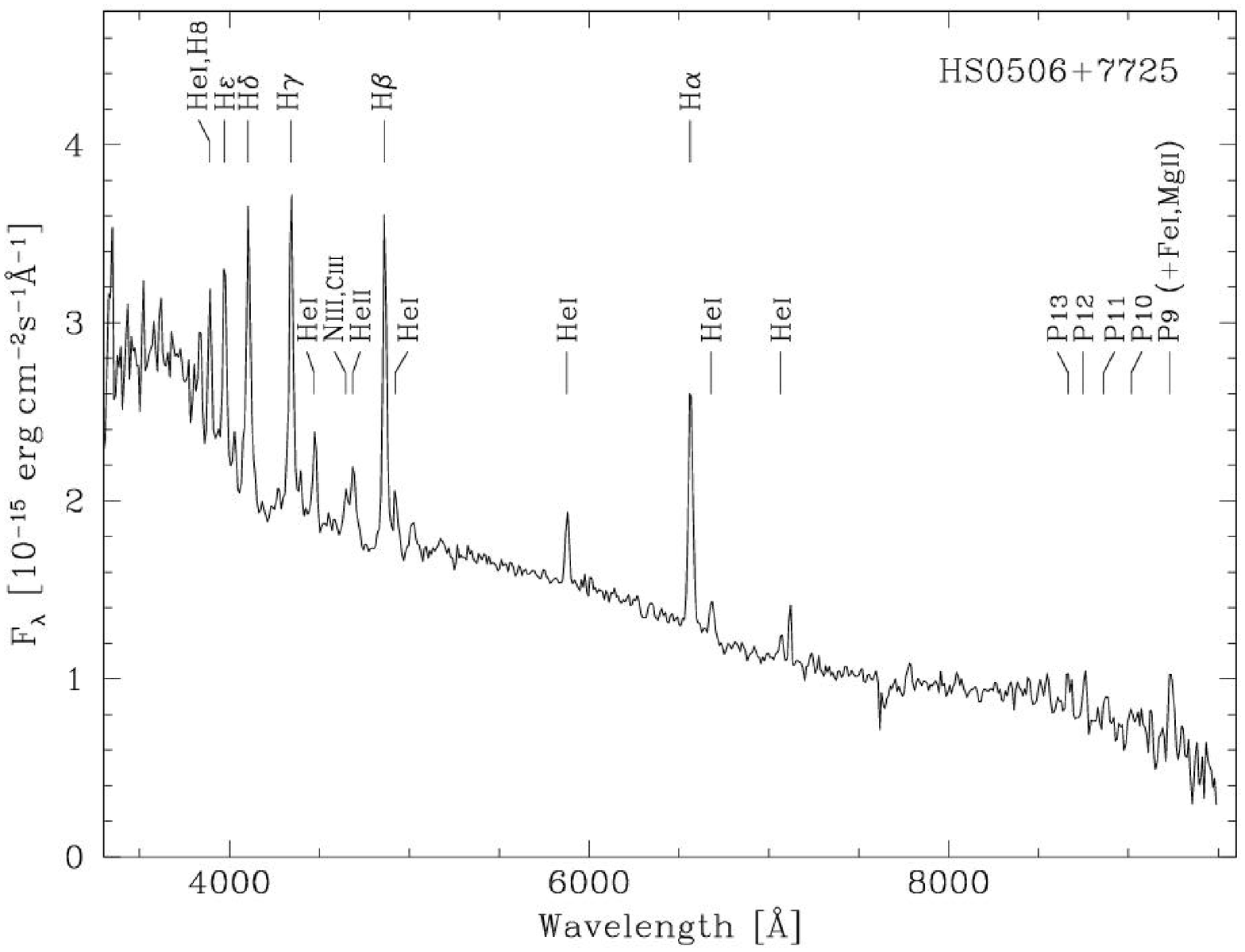}}
\caption[]{\label{f-idspec_hs0506} Identification spectrum of
  HS\,0506+7725 obtained at the Calar Alto 2.2\,m telescope on 1998
  February 2.}
\end{minipage}
\hfill
\begin{minipage}[t]{\columnwidth}
\centerline{\includegraphics[angle=-90,width=\columnwidth]{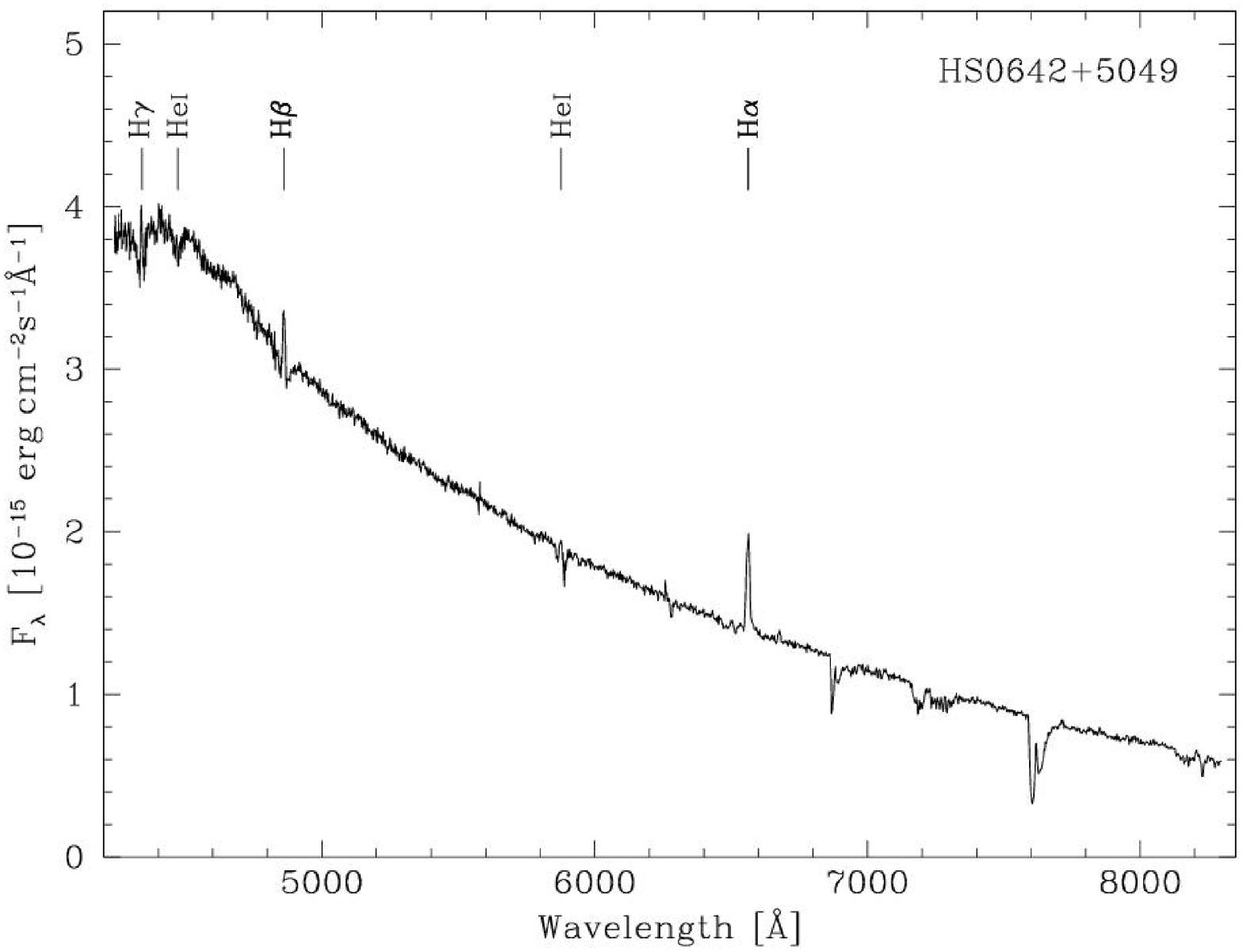}}
\caption[]{\label{f-averagespec_hs0642} Average of the 17 CAFOS G-100 
  spectra of
  HS\,0642+5049 obtained at the Calar Alto 2.2\,m telescope on 2004
  October 24.}
\end{minipage}
\end{figure*}

An identification spectrum of HS\,0139+0559 (see finding chart in
Fig.\,\ref{f-fc}) was obtained in October 1989 at the Calar Alto
3.5\,m telescope using the Boller\&Chivens spectrograph equipped with
a 120\AA/mm grating as part of a program to find blue stars
\citep{heberetal91-1}. This spectrum (Fig.\,\ref{f-idspec_hs0139})
is characterised by a blue continuum with strong absorption lines of the Balmer
series as well as of \Line{He}{I}{4471}. The absorption profiles have
a rectangular shape with a full width at zero intensity (FWZI) of
$\sim3500$\,\kms. The cores of \Hb\ and \Hg\ broad absorptions show
weak emission lines. No \Line{He}{II}{4686} is
observed. Overall, the spectrum resembles that of a high-mass transfer
rate accretion disc seen at a moderately low inclination, e.g. a dwarf
nova in outburst or a novalike variable.

HS\,0229+8016 (Fig.\,\ref{f-fc}) was first spectroscopically observed
at the Calar Alto 2.2\,m telescope on 1982 August 8, using the Boller
\& Chivens spectrograph. The spectrum (Fig.\,\ref{f-idspec_hs0229},
top panel) shows a blue continuum with the Balmer jump in emission,
superimposed by moderately strong Balmer emission lines. The higher
Balmer line profiles show evidence for a P-Cygni like structure with
blue absorption wings increasing in strength for the higher members of
the series. \Line{He}{I}{4471} is observed in absorption, and an
emission line near 4630\,\AA\ is detected, that we tentatively
identify as the N/C Bowen blend emission. Rather unusual is, however,
the fact that \Line{He}{II}{4686} is not detected in emission along
with the Bowen blend. HS\,0229+8016 was observed again at the Calar
Alto 2.2\,m telescope on 1998 October 5 using the Calar Alto Faint
Object Spectrograph (CAFOS), on this occasion looking nearly identical
to HS\,0139+8016 (Fig.\,\ref{f-idspec_hs0229}, bottom panel). The
spectral characteristics and the variability clearly identify
HS\,0229+8016 as a CV.

An identification spectrum of HS\,0506+7725 (Fig.\,\ref{f-fc}) was
obtained on 1998 February 2 on the Calar Alto 2.2\,m telescope using
CAFOS with the B-400 grism. The spectrum (Fig.\,\ref{f-idspec_hs0506})
displays a blue continuum with the Balmer and Paschen jumps in
emission, plus emission lines of hydrogen and \Ion{He}{I}. The
$\lambda\lambda4630-4650$ N/C Bowen blend and the \Line{He}{II}{4686}
emission lines are also detected. The strength of these
high-ionisation lines is typical of novalike variables, magnetic CVs
or nova remnants.

HS\,0642+5049 (Fig.\,\ref{f-fc}) was spectroscopically identified as a
CV on 1999 March 7 with CAFOS at the Calar Alto 2.2\,m telescope. The
spectrum of HS\,0642+5049 (Fig.\,\ref{f-averagespec_hs0642}) contains
a blue continuum with moderately strong \Ha\ emission. The \Hb\
and \Hg\ emission is embedded in broad absorption troughs, and weak
\Line{He}{I}{4471} absorption is also detected. No \Line{He}{II}{4686}
and $\lambda\lambda4630-4650$ N/C Bowen blend are detected.

In order to determine the orbital periods of HS\,0139+0559,
HS\,0229+8016, HS\,0506+7725 and HS\,0642+5049 we obtained
time-resolved spectroscopy at Calar Alto Observatory and Roche de los
Muchachos Observatory throughout the period October 2002 to October
2004 (Table\,\ref{t-obslog}). At the Calar Alto 2.2\,m telescope, we
used the CAFOS spectrograph equipped with the G-100 grating and a
$2\mathrm{k}\times2\mathrm{k}$ pixel SITe CCD. This setup, in conjunction
with a 1.2\,\arcsec slit, provided a spectral resolution of
$\sim4.1$\,\AA\ (full width at half maximum, FWHM) covering the range
4240--8300\,\AA. We used the double-armed TWIN spectrograph at the
Calar Alto 3.5\,m telescope equipped with the T05 grating in the blue
and the T06 grating in the red, providing a spectral resolution of
$\sim1.2$\,\AA\ (FWHM) in the ranges $\lambda\lambda3810-4940$ and
$\lambda\lambda6440-7510$. At the 2.5\,m Isaac Newton Telescope (INT),
we used the Intermediate Dispersion Spectrograph (IDS) equipped with
the R632V grating and the $2\mathrm{k}\times4\mathrm{k}$ pixel EEV10a
detector. Using a slit width of 1.5\,\arcsec this setup provided a
useful wavelenght of $\sim4400\,-\,6800$\,\AA\ and a spectral
resolution of $\sim2.3$\,\AA. Arc calibration spectra were interleaved
with the target observations every $\sim40$\,min.

In total, we obtained 55 spectra of HS\,0139+0559, 74 spectra of
HS\,0229+8016, 86 spectra of HS\,0506+7725 and 87 spectra of
HS\,0642+5049 (Table\,\ref{t-obslog}). The reduction of the follow-up
spectroscopy consisting of bias and flat-field corrections and optimal
extraction \citep{horne86-1} was carried out in
\texttt{IRAF}.\footnote{\texttt{IRAF} is distributed by the National
Optical Astronomy Observatories.} The wavelength calibration of the
extracted spectra was performed in \texttt{MOLLY}. The dispersion
relation was obtained by fitting a low-order polynomial to the arc
lines, with the RMS being less than one tenth of the dispersion in all
cases. The flexure of the telescope was accounted for by interpolating
between the two arc exposures bracketing the target spectra.

The time-resolved spectra of HS\,0139+0559, HS\,0229+8016 are similar
to the identification spectra shown in Fig.\,\ref{f-idspec_hs0139} and
Fig.\,\ref{f-idspec_hs0229} (bottom panel), with \Hb\ and \Hg\ in
absorption for most of the time, and occasionally showing signs of
emission cores. However, the G-100 spectra cover \Ha\ as well, which
is observed in emission throughout, with low equivalent widths in the
range $3-6$\,\AA. The time-resolved spectra of HS\,0506+7725 are very
similar to the identification spectrum shown in
Fig.\,\ref{f-idspec_hs0506}, with difference of having no
\Line{He}{II}{4686} emission. The \Ha\ emission line has an average
equivalent width of $\sim15$\,\AA. The time-resolved spectra of
HS\,0642+5049 show \Ha\ in emission with an average equivalent width
of $\sim5$\,\AA, and weak \Hb\ and \Hg\ emission embedded in broad
absorption lines (Fig.\,\ref{f-averagespec_hs0642}).

\begin{table*}[t]
\setlength{\tabcolsep}{1.1ex}
\caption[]{Log of the observations\label{t-obslog}.}
\vspace*{-2.5ex}

\begin{minipage}[t]{8.8cm}
\begin{tabular}[t]{lccccc}
\hline\noalign{\smallskip}
Date & UT Time &  Filter/ & Exp. & Frames & Mean     \\    
     &         &  Grating & (s)  &        & mag.     \\    
\hline\noalign{\smallskip}
\multicolumn{6}{c}{\textbf{HS\,0139+0559}} \\
\multicolumn{6}{l}{\textbf{Spectroscopy: Calar Alto 3.5\,m \& B\&C}} \\
1989 Jan 22 & 19:21       & 120\AA/mm    & 900 & 1 & -\\
\multicolumn{6}{l}{\textbf{Spectroscopy: Calar Alto 3.5\,m \& TWIN}} \\
2002 Oct 28 & 23:53-00:03 & T08/T01      & 300 & 1 & - \\
2002 Oct 29 & 22:36-03:21 & T05/T06      & 600 & 16 & - \\
\multicolumn{6}{l}{\textbf{Spectroscopy: Calar Alto 2.2\,m \& CAFOS}} \\
2003 Dec 24 & 18:25-00:10 & G-100 & 600 & 27 & $\simeq15.4$ \\
2003 Dec 26 & 18:34-20:48 & G-100 & 600 & 12 & $\simeq15.1$ \\
\multicolumn{6}{l}{\textbf{Spectroscopy: Hamburg-Schmidt telescope at Calar 
Alto}} \\
1981 Nov 01 & - & Prism & 3600 & 1 & $\simeq14.9$ \\
1981 Nov 02 & - & Prism & 3600 & 1 & $\simeq14.9$ \\
\multicolumn{6}{l}{\textbf{Photometry: Braeside Observatory}} \\
1999 Dec 30 & 02:15-07:04 & $R$ & 70 & 293 & - \\ 
2000 Jan 04 & 02:31-06:04 & $B$ & 50 & 136 & - \\ 
\noalign{\smallskip}
\multicolumn {6}{c}{\textbf{HS\,0229+8016}} \\
\multicolumn{6}{l}{\textbf{Spectroscopy: Calar Alto 2.2\,m \& B\&C}} \\
1992 Aug 08 & 03:16       &       & 900 & 1 & $\simeq15.0$ \\
\multicolumn{6}{l}{\textbf{Spectroscopy: Calar Alto 2.2\,m \& CAFOS}} \\
1998 Oct 05 & 01:37       & B-200 & 1800 & 1 & $\simeq13.7$ \\
2003 Dec 14 & 19:18-00:55 & G-100 & 600 & 24 & $\simeq14.2$ \\
2003 Dec 16 & 20:21-03:08 & G-100 & 600 & 27 & - \\ 
2003 Dec 23 & 18:49-21:01 & G-100 & 600 & 12 & $\simeq14.5$ \\
2003 Dec 26 & 21:34-22:53 & G-100 & 600 & 7  & $\simeq14.1$ \\
2003 Dec 27 & 18:46-19:22 & G-100 & 600 & 4  & $\simeq13.7$ \\
\multicolumn{6}{l}{\textbf{Spectroscopy: Hamburg-Schmidt telescope at Calar 
Alto}} \\
1986 Nov 05 & - & Prism & 3600 & 1 & $\simeq14.4$ \\
1986 Nov 06 & - & Prism & 3600 & 1 & $\simeq14.2$ \\
1994 Nov 12 & - & Prism & 3600 & 1 & $\simeq14.6$ \\
\multicolumn{6}{l}{\textbf{Photometry: Kryoneri Observatory}} \\
2002 Sep 20 & 01:30-03:28 & $R$ & 5 & 840 & $\simeq14.2$ \\ 
2002 Sep 20 & 22:37-03:27 & $R$ & 5 & 1680 & $\simeq14.3$ \\ 
\multicolumn{6}{l}{\textbf{Photometry: Tuorla Observatory}} \\
2003 Jan 10 & 15:41-23:45 & Clear & 30 & 695 & $\simeq14.3$ \\
\noalign{\smallskip}
\multicolumn{6}{c}{\textbf{HS\,0506+7725}}\\
\multicolumn{6}{l}{\textbf{Spectroscopy: Calar Alto 3.5\,m \& TWIN}} \\
2002 Dec 03 & 03:22-05:09 & T05/T06 & 600 & 10 & - \\
\multicolumn{6}{l}{\textbf{Spectroscopy: Calar Alto 2.2\,m \& CAFOS}} \\
1998 Feb 02 & 19:43       & B-400 & 180 &  1 & -  \\
2002 Dec 09 & 00:19-00:30 & G-100 & 600 &  2 & $\simeq15.4$ \\
2003 Dec 13 & 23:25-03:53 & G-100 & 600 &  23 & $\simeq14.7$ \\
2003 Dec 15 & 01:34-06:13 & G-100 & 600 &  23 & $\simeq14.9$ \\
2003 Dec 17 & 03:40-05:57 & G-100 & 600 &  12 & - \\
2003 Dec 23 & 22:19-00:31 & G-100 & 600 &  12 & $\simeq14.9$ \\
2003 Dec 26 & 02:52-03:28 & G-100 & 600 &  4 & $\simeq14.9$ \\
\noalign{\smallskip}\hline
\end{tabular}
\end{minipage}
\hfill
\begin{minipage}[t]{8.8cm}
\begin{tabular}[t]{lccccc}
\hline\noalign{\smallskip}
Date & UT Time &  Filter/ & Exp. & Frames & Mean     \\    
     &         &  Grating & (s)  &        & mag.     \\    
\hline\noalign{\smallskip}
\multicolumn{6}{l}{\textbf{Spectroscopy: Hamburg-Schmidt telescope at Calar 
Alto}} \\
1987 Oct 29 & - & Prism & 3600 & 1 & $\simeq15.1$ \\
1994 Jan 12 & - & Prism & 3600 & 1 & $\simeq15.3$ \\
1995 Oct 23 & - & Prism & 3600 & 1 & $\simeq18.3$ \\
\multicolumn{6}{l}{\textbf{Photometry: Kryoneri Observatory}} \\
2002 Oct 08 & 22:46-01:25 & $R$ & 10 & 599 & $\simeq15.2$ \\ 
\multicolumn{6}{l}{\textbf{Photometry: Tuorla Observatory}} \\
2003 Jan 04 & 16:13-20:16 & Clear & 120 & 32 & $\simeq15.7$ \\
2003 Jan 06 & 15:47-20:46 & Clear & 60 & 213 & $\simeq15.8$ \\
2003 Jan 16 & 20:18-22:40 & Clear & 60 & 115 & $\simeq15.9$ \\
\multicolumn{6}{l}{\textbf{Photometry: OGS}} \\
2003 Nov 15 & 03:21-06:37 & Clear & 5  & 1110 & $\simeq14.6$ \\
\multicolumn{6}{l}{\textbf{Photometry: Calar Alto 2.2\,m \& CAFOS}} \\
2003 Jan 01 & 01:01-03:24 & $V$ & 30 & 85 & $\simeq15.1$ \\
2003 Dec 16 & 00:05-05:11 & Clear & 15-30 & 346 & $\simeq14.8$ \\
2003 Dec 25 & 19:15-02:29 & Clear & 15-30 & 296 & $\simeq14.9$ \\
2003 Dec 27 & 04:03-06:16 & Clear & 20 & 149 & $\simeq14.9$ \\
2003 Dec 27 & 20:33-23:55 & Clear & 20 & 288 & $\simeq14.8$ \\
\noalign{\medskip}
\multicolumn{6}{c}{\textbf{HS\,0642+5049}}\\
\multicolumn{6}{l}{\textbf{Spectroscopy: Calar Alto 2.2\,m \& CAFOS}} \\
1999 Mar 07 & 19:11       & B-400 & 300 & 1 & - \\
2003 Apr 27 & 20:03-21:53 & G-100 & 600 & 9 & $\simeq15.6$ \\
2003 May 10 & 20:20-21:21 & G-100 & 600 & 6 & - \\
2003 May 11 & 20:17-21:20 & G-100 & 600 & 6 & - \\
2003 Dec 25 & 00:56-05:35 & G-100 & 600 & 24 & $\simeq15.5$ \\
2003 Dec 27 & 02:50-03:01 & G-100 & 600 & 2 & $\simeq15.3$ \\
2004 Oct 24 & 02:12-05:26 & G-100 & 600 & 17 & $\simeq15.3$ \\
2004 Oct 26 & 04:25-05:31 & G-100 & 600 & 7 & $\simeq15.4$ \\
\multicolumn{6}{l}{\textbf{Spectroscopy: INT 2.5\,m \& IDS}} \\
2003 Apr 25 & 21:21-22:14 & R632V & 600 & 6 & - \\
2003 Apr 26 & 21:17-21:49 & R632V & 600-900 & 4 & - \\
2003 Apr 28 & 21:15-22:08 & R632V & 600 & 6 & - \\
\multicolumn{6}{l}{\textbf{Spectroscopy: Hamburg-Schmidt telescope at Calar 
Alto}} \\
1991 Nov 10 & - & Prism & 3600 & 1 & $\simeq15.8$ \\
1993 Oct 24 & - & Prism & 3600 & 1 & $\simeq16.0$ \\
\multicolumn{6}{l}{\textbf{Photometry: Tuorla Observatory}} \\
2003 Dec 30 & 19:58-21:06 & Clear & 120 & 46 & $\simeq15.4$ \\
2004 Jan 01 & 18:37-23:27 & Clear & 60 & 250 & $\simeq15.5$ \\
\multicolumn{6}{l}{\textbf{Photometry: Calar Alto 2.2\,m \& CAFOS}} \\ 
2003 Dec 26 & 04:20-06:14 & Clear & 20-30 & 149 & $\simeq15.6$ \\
2004 Oct 22 & 04:01-05:25 & Clear & 15 & 59 & $\simeq15.7$ \\
2004 Oct 25 & 01:42-05:51 & Clear & 10-15 & 463 & $\simeq15.5$ \\
\multicolumn{6}{l}{\textbf{Photometry: IAC80}} \\ 
2004 Dec 02 & 04:01-05:00 & Clear & 10 & 150 & $\simeq15.4$ \\
2004 Dec 07 & 05:10-06:11 & Clear & 15 & 139 & $\simeq15.3$ \\
2004 Dec 08 & 01:00-04:41 & Clear & 10 & 528 & $\simeq15.3$ \\
2004 Dec 09 & 00:13-05:14 & Clear & 10 & 738 & $\simeq15.3$ \\
\noalign{\smallskip}\hline
\end{tabular}
\end{minipage}
\end{table*}

\subsection{Photometry}
The spectroscopy of the four new cataclysmic variables was
supplemented by differential CCD photometry obtained at six
different telescopes (Table\,\ref{t-obslog}). Brief descriptions of
the used instrumentation and the data reduction techniques are given
below. 

\paragraph{Braeside Observatory.}  
Differential $B$ and $R$ photometry of HS\,0139+0559 was obtained in
December 1999/January 2000 at Braeside Observatory using the 0.41\,m
reflector together with a SITe $512\times512$ pixel CCD
camera. The raw images were bias-subtracted, dark current-subtracted
and flat-fielded in a standard manner.

\paragraph{Kryoneri Observatory.}
We obtained differential $R$-band photometry of HS\,0229+8016 (2002
September 19 \& 20) and HS\,0506+7725 (2002 October 8) at Kryoneri
Observatory using the 1.2\,m telescope equipped with a Photometrics
SI-502 $516\times516$ pixel CCD camera. The data were reduced
using the pipeline discribed in \citet{gaensickeetal04-1}, which
pre-processes the raw images in \texttt{MIDAS} and extracts aperture
photometry using the \texttt{Sextractor}
\citep{bertin+arnouts96-1}. The instrumental magnitudes of
HS\,0229+8016 were converted into apparent magntitudes using the
comparison star labelled 'C1' in Fig.\,\ref{f-fc} (USNO-A2.0
1650-00512682, $R=12.6$). We found a mean magnitude of $R\simeq14.3$
during both nights. The apparent magnitudes of HS\,0506+7725 were computed
using the comparison star 'C2' (USNO-A2.0 1650-00942250, $R=13.1$),
and an average magnitude of $R\simeq15.2$ was found.

\paragraph{Tuorla Observatory.}
We used the 0.7\,m Schmidt-Vaisala telescope at Tuorla Observatory,
equipped with a SBIG ST-8 CCD camera to obtain filterless photometry
of HS\,0229+8016 (2003 January 10), HS\,0506+7725 (2003 January 4, 6
\& 16) and HS\,0642+5049 (2003 December 30 \& 2004 January 1). The
reduction of the observations was carried out in the same way as
described for the Kryoneri data. Using the same comparison stars as
above, we found the mean magnitudes of HS\,0229+8016 and HS\,0506+7725
to be $\simeq14.3$ and $\simeq15.8$, respectively. The apparent
magnitudes of HS\,0642+5049 were extracted using the comparison star
'C3'in Fig.\,\ref{f-fc} (USNO-A2.0 1350-06806656, $R=12.4$), and gave a
mean magnitude of $\simeq15.5$.

\paragraph{Observatorio del Teide.}
At the Observatorio del Teide on Tenerife we used the 1\,m Optical
Ground Station (OGS) and the 0.82\,m IAC80 telescopes to obtain CCD
photometry of HS\,0506+7725 and HS\,0642+5049. The telescopes were
equipped with Thomson $1\mathrm{k}\times1\mathrm{k}$ pixel CCD cameras. On 2003
November 15, the OGS was used to obtain differential photometry of
HS\,0506+7225 in white light. The data were taken using $2\times2$
binning and windowing to improve the time resolution. The images were
bias and flat-field corrected and aligned within
\texttt{IRAF}. Instrumental magnitudes of the CV and the comparison
star 'C2' were then extracted using the point spread function (PSF)
photometry tasks package within \texttt{IRAF}. We found the mean
magnitude of HS\,0506+7725 during that night to be $\simeq14.6$. On
2004 December 1, 6, 7 \& 8, we used the IAC80 telescope to obtain
filterless photometry of HS\,0642+5049. In order to achieve a high
time resolution, windowing and binning $2\times2$ were applied. The
data were reduced using \texttt{MIDAS} in the same manner as described
for the Kryoneri run. The apparent magnitudes were extracted using the
comparison star 'C3', resulting in a mean magnitude of $\simeq$ 15.3.

\paragraph{Calar Alto Observatory.}
During January 2003 and October 2004, we used CAFOS with the SITe
$2\mathrm{k}\times2\mathrm{k}$ pixel CCD camera on the 2.2\,m
telescope to obtain filterless differential photometry of
HS\,0506+7725 and HS\,0642+5049 when the atmospheric conditions were
too poor for spectroscopy. Only a small part of the CCD was read out
in order to improve the time resolution.  The data were reduced in an
analoguous fashion as described for the Kryoneri observations
above. The mean magnitudes of HS\,0506+7725 and HS\,0642+5049 were
found to be $\simeq$14.9 and $\simeq$15.6 respectively.

\begin{figure*}
\begin{minipage}[t]{\columnwidth}
\centerline{\includegraphics[angle=-90,width=\columnwidth]{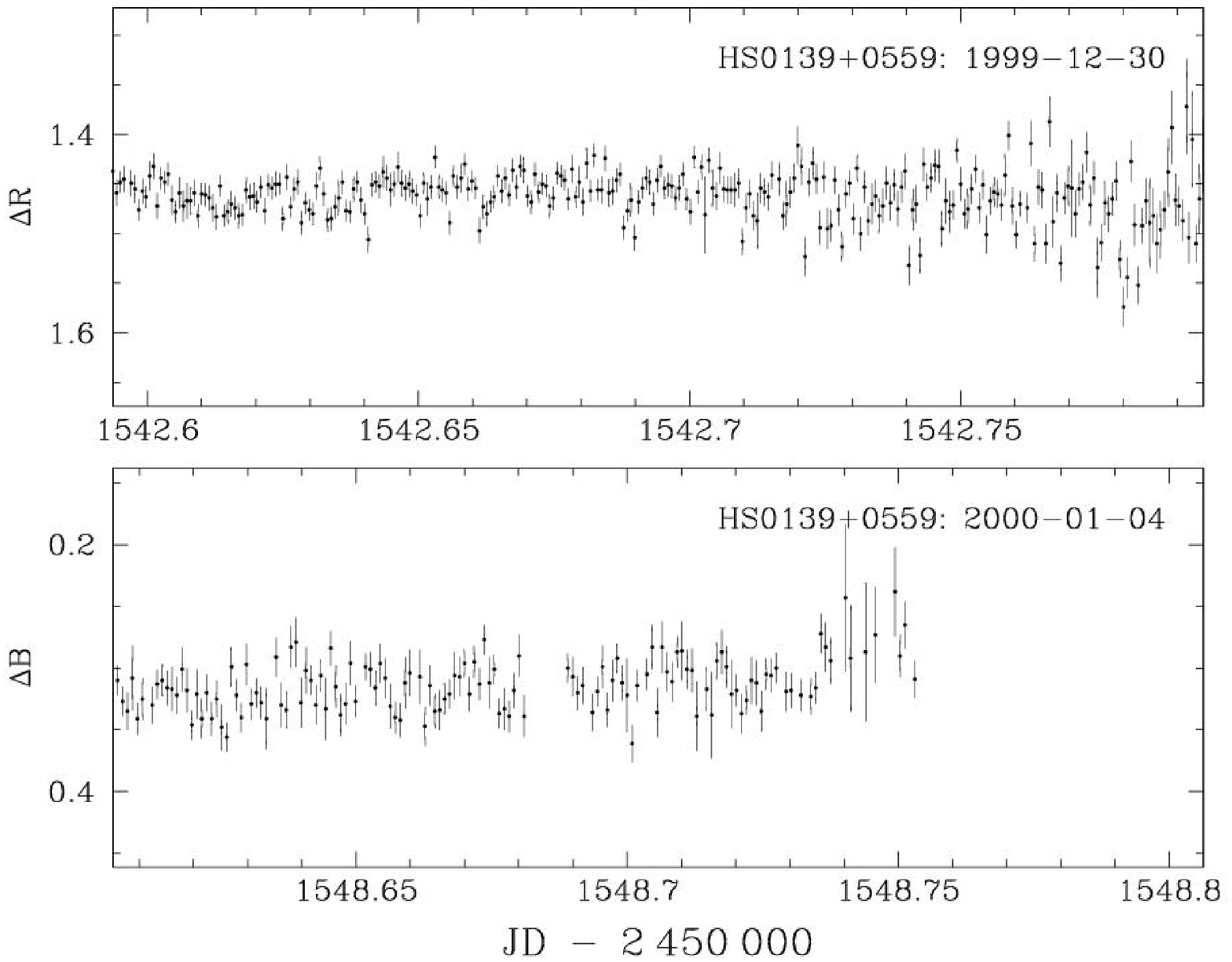}}
\caption[]{\label{f-lc_hs0139} Differential CCD $R$-band (top panel)
and $B$-band (bottom panel) photometry of HS\,0139+0559 obtained at
the Braeside observatory.}
\end{minipage}
\hfill
\begin{minipage}[t]{\columnwidth}
\centerline{\includegraphics[angle=-90,width=\columnwidth]{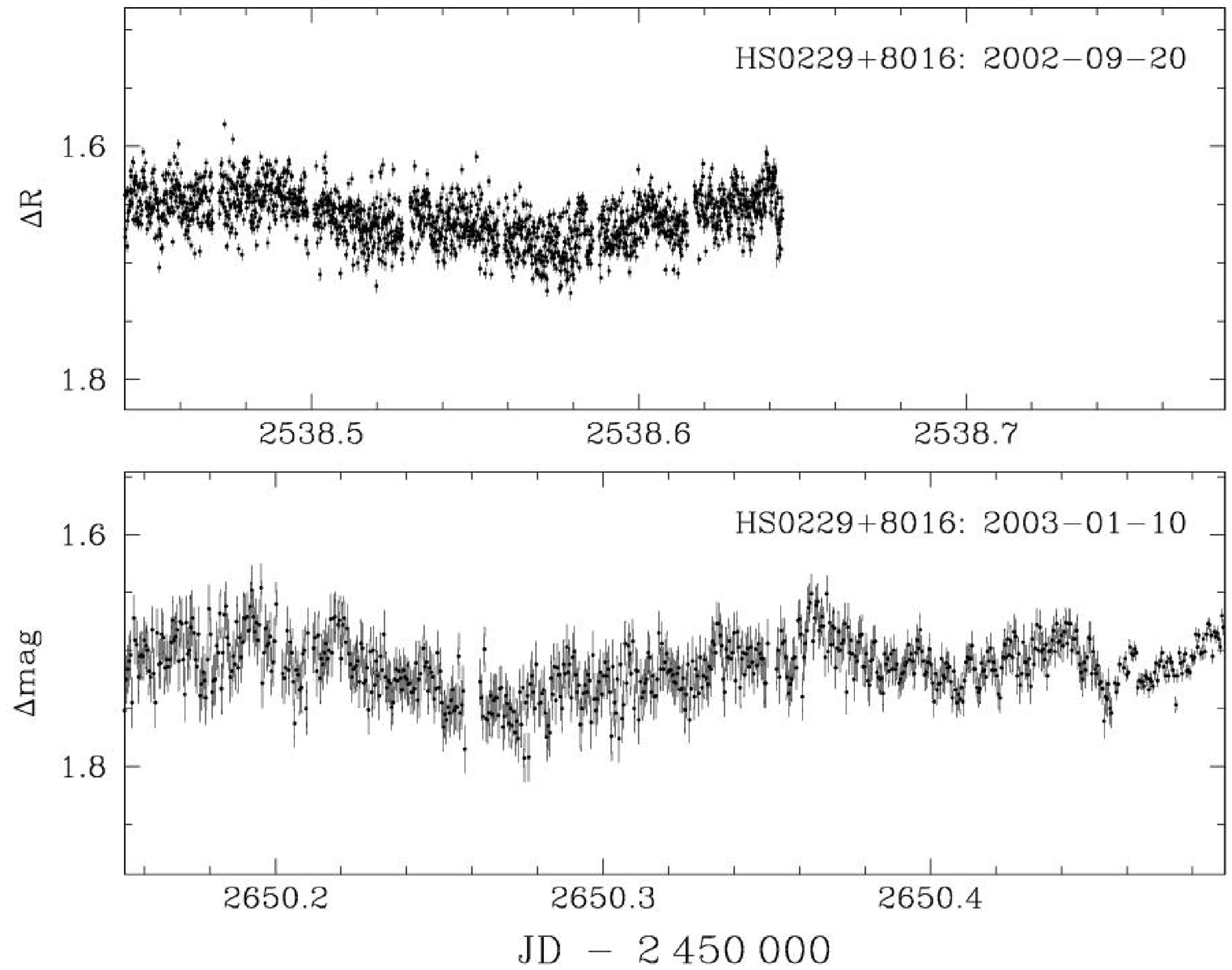}}
\caption[]{\label{f-lc_hs0229} Sample light curves of
HS\,0229+8016. Top panel: $R$-band data obtained at the Kryoneri
observatory. Bottom panel: filterless data  obtained at the Tuorla observatory.}
\end{minipage}

\medskip
\begin{minipage}[t]{\columnwidth}
\centerline{\includegraphics[angle=-90,width=\columnwidth]{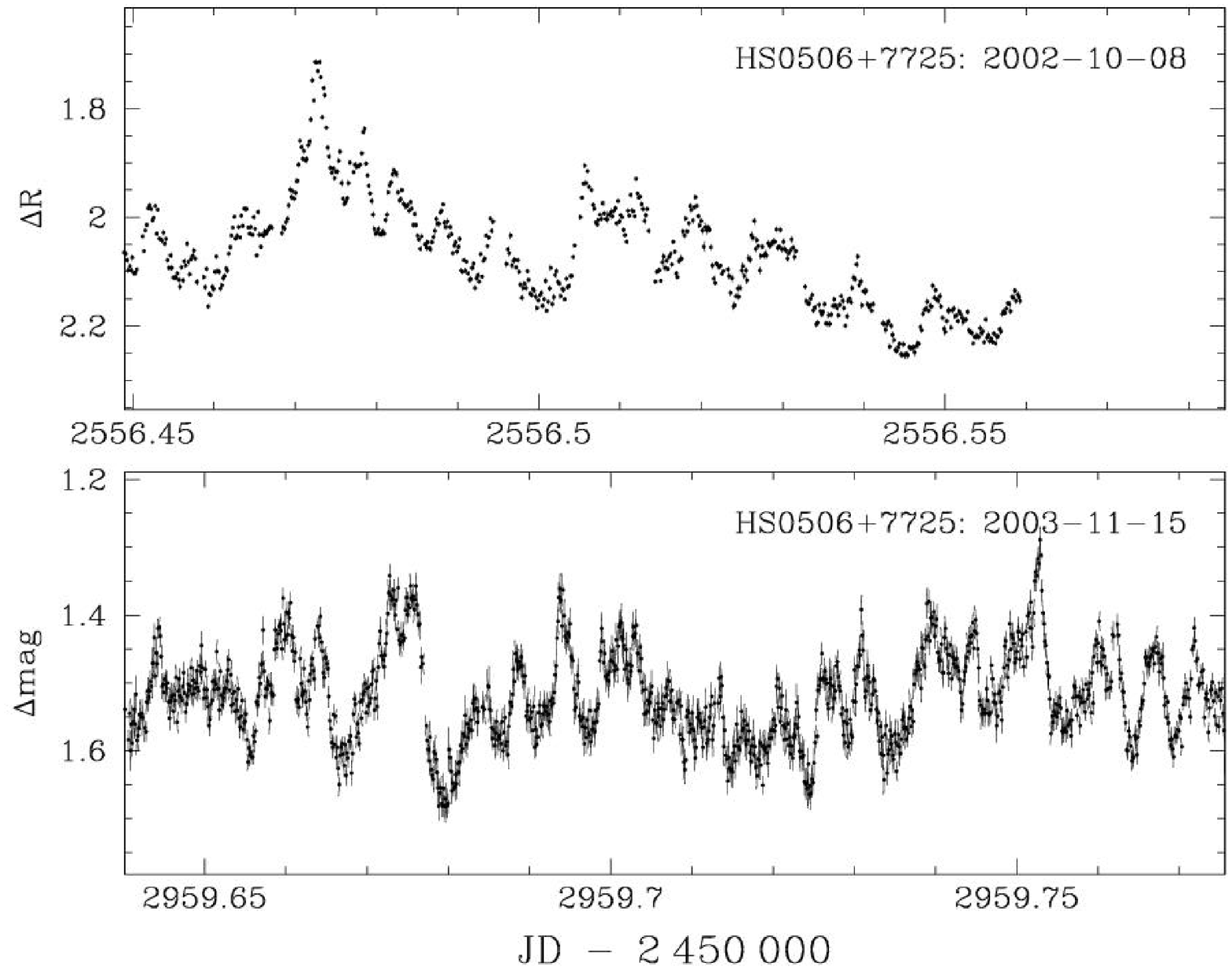}}
\caption[]{\label{f-lc_hs0506} Sample light curves of
HS\,0506+7725. Top panel: $R$-band data obtained at the Kryoneri
observatory. Bottom panel: filterless data  obtained at the OGS telescope.}
\end{minipage}
\hfill
\begin{minipage}[t]{\columnwidth}
\centerline{\includegraphics[angle=-90,width=\columnwidth]{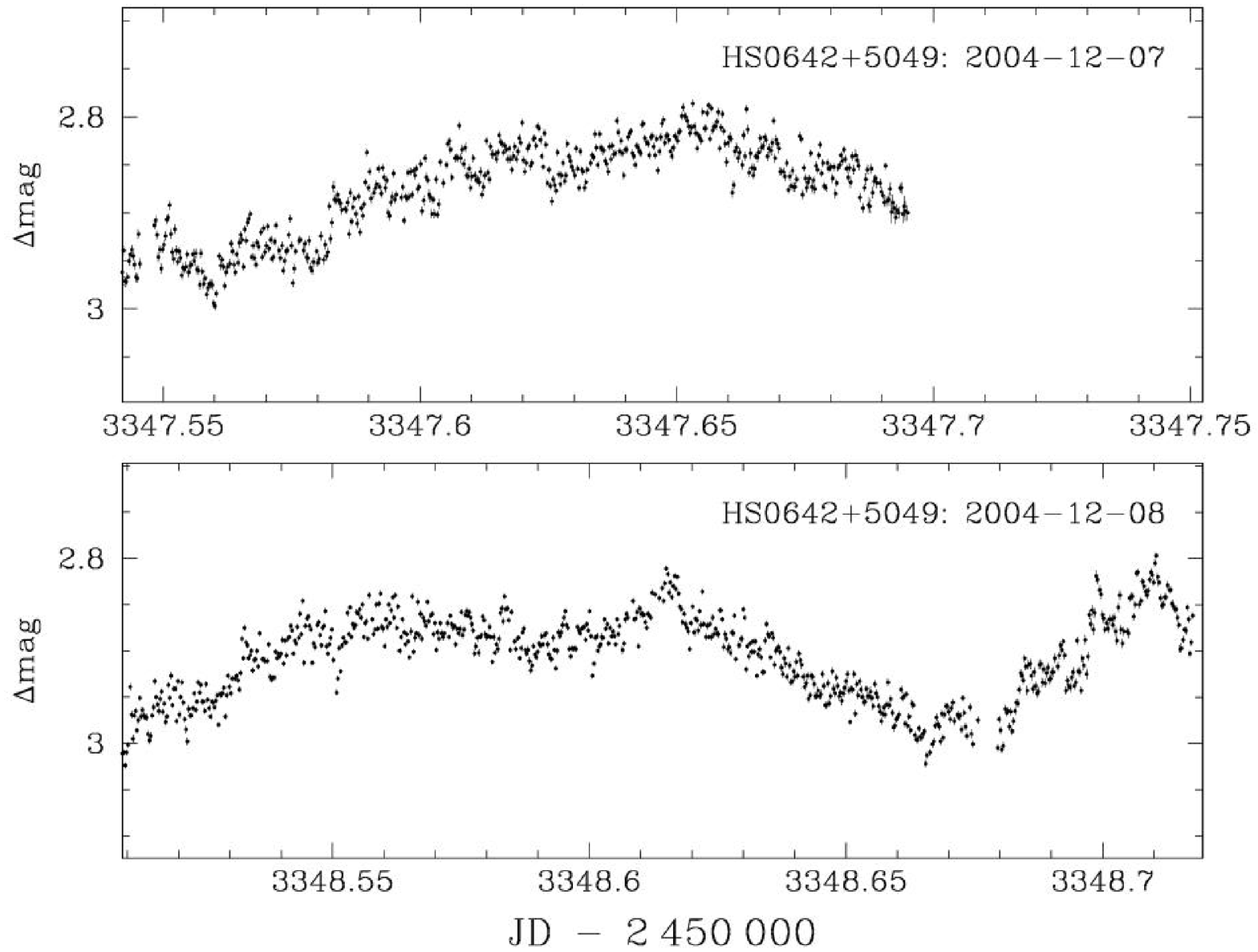}}
\caption[]{\label{f-lc_hs0642} Sample filterless light curves of HS\,0642+5049
obtained at the IAC80 telescope.}
\end{minipage}
\end{figure*}

\paragraph{Light curve morphology.}
The light curves of HS\,0139+0559 (Fig.\,\ref{f-lc_hs0139}) and
HS\,0229+8016 (Fig.\,\ref{f-lc_hs0229}) display very little
variability on nightly time scales, with amplitudes $\la0.02$\,mag and
$\sim0.05$\,mag, respectively.  In the case of HS\,0229+8016, a
low-amplitude modulation with a period of $\sim4$\,h is consistently
detected during the two longest observations. HS\,0506+7725 exhibits
short-period variability with an amplitude of $\sim0.2-0.4$\,mag which
appear to be quasi-periodic oscillations on time scales of
$\sim10-20$\,min (Fig.\,\ref{f-lc_hs0506}). No clearly repeating
variation is detected on time scales of several hours (i.e. a putative
orbital modulation). The light curves of HS\,0642+5049 from the IAC80
(Fig.\,\ref{f-lc_hs0642}) show a modulation with a period of 
$\sim3.5$\,hr which we interpret as the orbital period of the
system. No substantial flickering activity is detected.

The USNO-A2.0 catalogue lists HS\,0139+0559 with $B=R=14.4$ and we
found $V\simeq15.4$ and $B\simeq14.9$ in our
observations. HS\,0229+8016 has $B=13.9$ and $R=13.8$ in the USNO-A2.0
catalogue, and has been found during our observations mostly near
$B\simeq V\simeq14.0-14.6$, except on one occasion (August 1992) when
it was as faint as $V\simeq15.0$\,mag. HS\,0506+7725 is listed with
$B=15.3$ and $R=15.6$ in the USNO-A2.0 catalogue, and our data
provides evidence for long-term variability of the mean magnitude by
$\sim1$\,mag, but dropping on one occasion (October 1995) into a deep
low state at $B\simeq18.3$. HS\,0642+5049 is found in the USNO-A2.0
catalogue with $B=16.6$ and $R=16.9$ and we found
$V\simeq15.3-15.5$\,mag without significant long-term variability of
the system.

\section{Analysis}

\subsection{\label{s-ana0139} HS\,0139+0559}
\begin{figure*}
\begin{minipage}[t]{\columnwidth}
\centerline{\includegraphics[angle=-90,width=8.8cm]{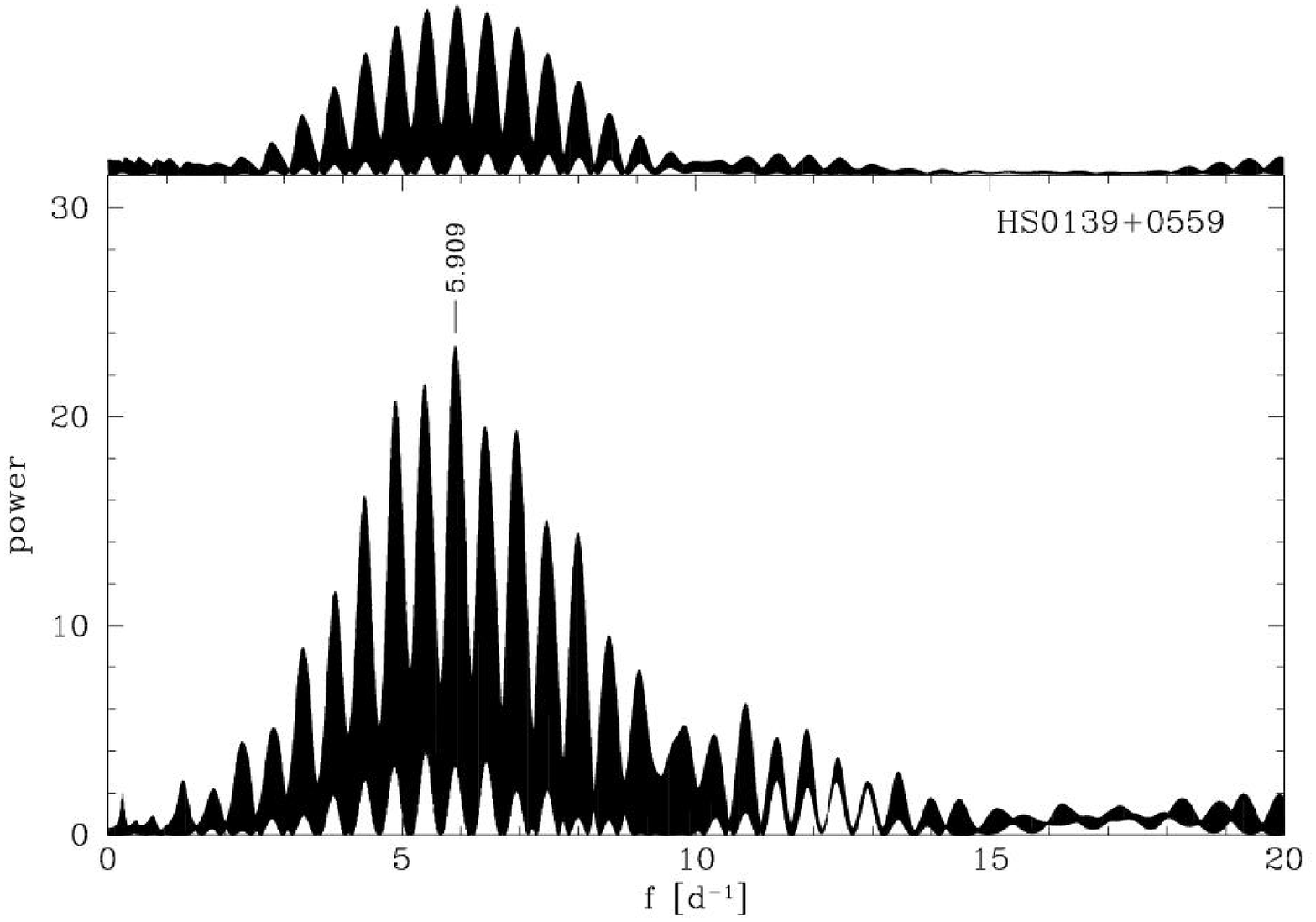}}
\caption[]{\label{f-scarglehs0139} The Scargle periodogram of the
radial velocities of HS\,0139+0559 measured from the $\Ha$ emission
line. The periodogram from a set of fake radial velocities is
shown in the top panel. }
\end{minipage}
\hfill
\begin{minipage}[t]{\columnwidth}
\centerline{\includegraphics[angle=-90,width=8.8cm]{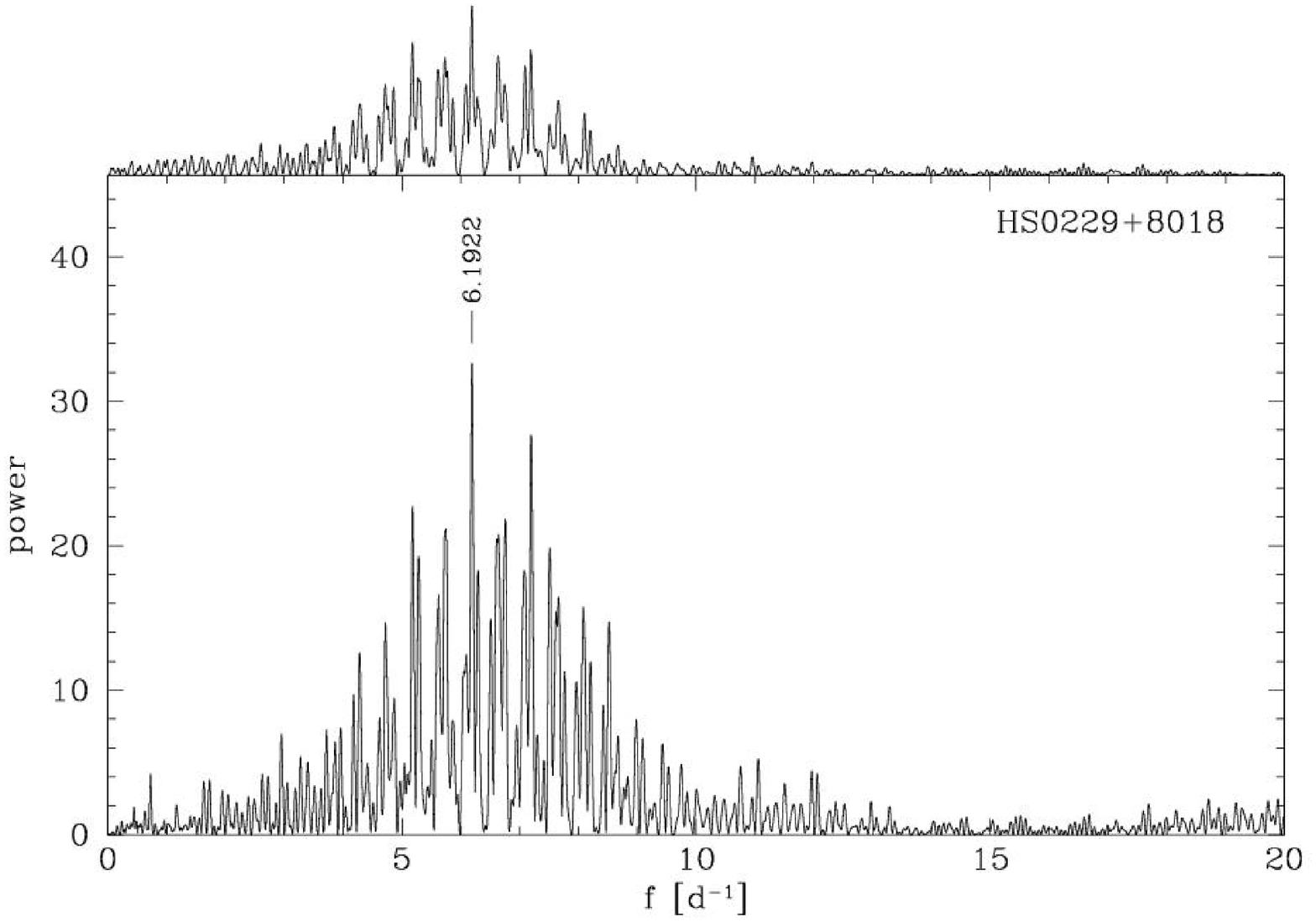}}
\caption[]{\label{f-scarglehs0229} The Scargle periodogram of the
radial velocities of HS\,0229+8016 measured from the $\Ha$ emission line. The
periodogram from a set of fake radial velocities is shown in the top panel. }
\end{minipage}

\medskip
\begin{minipage}[t]{\columnwidth}
\centerline{\includegraphics[angle=-90,width=8.8cm]{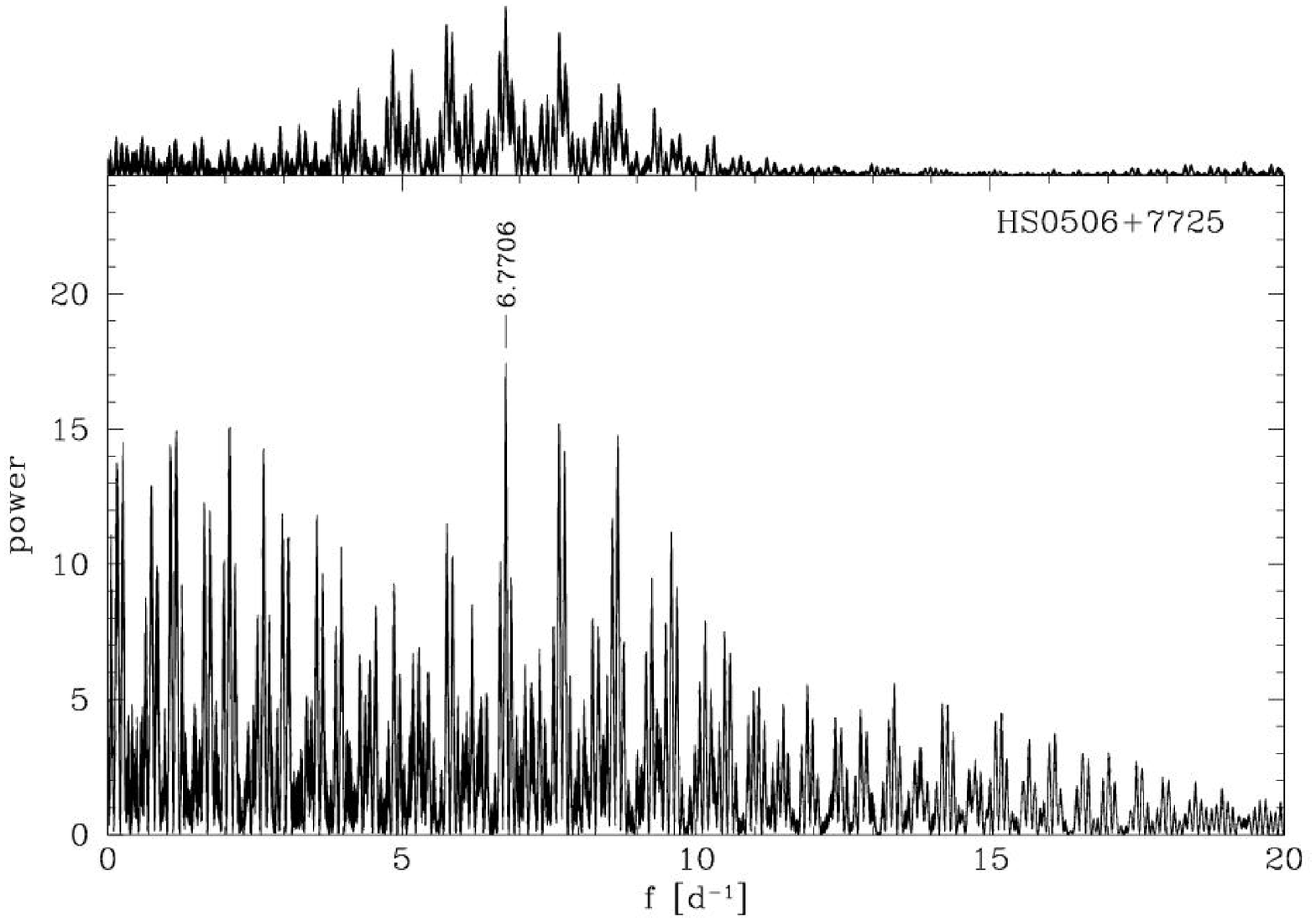}}
\caption[]{\label{f-scarglehs0506} The Scargle periodogram of the
radial velocities of HS\,0506+7725 measured from the $\Ha$ emission
line by using the double Gaussian method. The periodogram from a set
of fake radial velocities is shown in the top panel. }
\end{minipage}
\hfill
\begin{minipage}[t]{\columnwidth}
\centerline{\includegraphics[angle=-90,width=8.8cm]{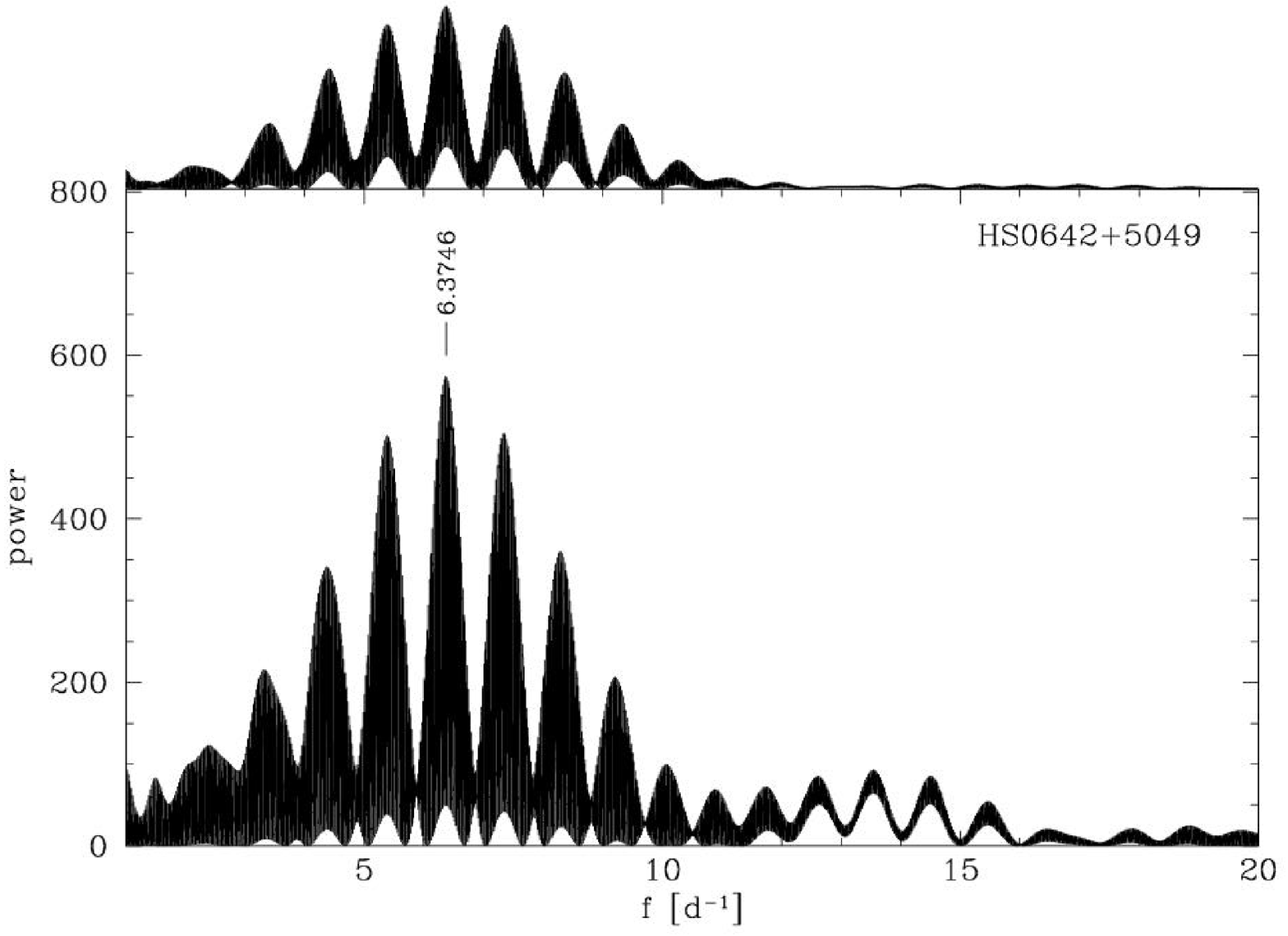}}
\caption[]{\label{f-scarglehs0642_phot} The Scargle periodogram of
HS\,0642+5049 computed from the three longest nights of differential
photometry obtained at the Calar Alto (2004 October 25) and the IAC80
(2004 December 8 \& 9). The periodogram from a fake data set assuming a
period of 225.90 min is shown in the top panel}
\end{minipage}
\end{figure*}

In order to determine the orbital period of HS\,0139+0559 we have
measured the radial velocity variations of \Ha\
by convolving the observed line profiles with a single Gaussion of
FWHM=600\,\kms. The spectra were continuum-normalised prior to this
analysis. A \citet{scargle82-1} period analysis of the radial velocity
measurements was performed using the \texttt{MIDAS/TSA} context. The
resulting periodogram (Fig.\,\ref{f-scarglehs0139}) shows a strong
signal at $5.909\pm0.012$\,\id (where the error is determined from
fitting a sine wave to the radial velocity variation, see
Table\,\ref{t-rvfits} for the full fit parameter), surrounded by
one-day aliases. In order to test the significance of the detected
signal, we have created a set of fake radial velocities by evaluating
a sine function with a frequency of $5.909$\,\id\ at the exact times
of the observed spectroscopic data. The amplitude of the sine wave was
adjusted to reflect the observed radial velocity amplitude, and the
fake radial velocity measurements were randomly offset from the
computed sine wave using the observed errors. The periodogram of the
fake data reproduces the alias structure of the periodogram computed
from the observations very well. We conclude that the orbital period
of HS\,0139+0559 is $\Porb=243.69\pm0.49$\,min. Folding the radial
velocity measurements over that period results in a quasi-sinusoidal
radial velocity curve with an amplitude of \textbf{$84.4\pm4.8$\,\kms}
(Fig.\,\ref{f-rvfolded}, top panel).

For completeness, we have also used the Braeside $B$ and $R$ band
photometry for a period analysis. As suggested by the flat light curve
(Fig.\,\ref{f-lc_hs0139}) no significant signal is detected in the
Scargle periodogram.

\subsection{HS\,0229+8016}
An analogous radial velocity analysis as described in the previous
section was carried out for HS\,0229+8016. Inspection of the Scargle
periodogram (Fig.\,\ref{f-scarglehs0229}) shows a somewhat more
complex alias structure as a result of the inhomogeneous spacing of
the spectroscopic observations. The strongest signal is found at
$6.1922\pm0.0013$\,\id\ (where the error is again computed from a sine
fit to the radial velocity data, Table\,\ref{t-rvfits}) and a fake
data set computed using this frequency reproduces well the overall
alias structure. We conclude that the most likely value for the
orbital period of HS\,0229+0559 is $\Porb=232.550\pm0.049$\,min. The
phase-folded radial velocity curve shows a quasi-sinusoidal modulation
with an amplitude of \textbf{$179.0\pm5.2$\,\kms} (Table\,\ref{t-rvfits}).

Scargle periodograms computed from the two longest photometry runs on
HS\,0229+0559 are dominated by a broad signal near 5.2\,\id\ (Kryoneri
data) and 6.2\,\id\ (Tuorla), which are consistent with the
spectroscopic period or its one-day alias. While our photometric data
is not sufficient to improve the period determination of
HS\,0229+0559, it suggests that the orbital period of HS\,0229+0559
can be refined by a sequence of sufficiently long photometric
time-series.

\subsection{HS\,0506+7725}
\begin{figure}
\centerline{\includegraphics[width=\columnwidth]{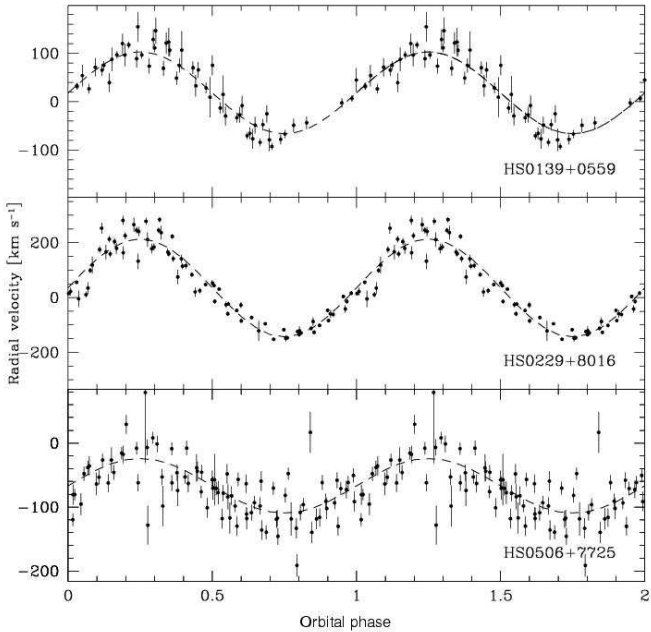}}
\caption[]{\label{f-rvfolded} $\Ha$ radial velocities of HS\,0139+0559
(top panel), HS\,0229+8016 (middle panel) and HS\,0506+7725 (bottom
panel) folded on the period of 243.69\,min, 232.550\,min and 212.7\,min
respectively. The dashed lines are the best sine fits to the folded
velocities. }
\end{figure}

Our initial analysis of the HS\,0506+7725 \Ha\ radial velocity
variation was carried again applying a single-Gaussian convolution on
the continuum-normalised line profiles. However, the Scargle
periodogram computed from the measured radial velocity variations
turned out to be dominated by a variety of signals in the range
$\sim1-5$\,\id, none of which resulted in a plausible phase-folded
radial velocity curve. In a second attempt, we applied the double
Gaussian method of \citet{schneider+young80-2}, using a Gaussian
FWHM=700\,$\kms$ and a separation of 1500\,$\kms$ which measures the
radial velocity variation of the line wings. The Scargle periodogram
resulting from these radial velocity measurements includes several
peaks in the range $5-9$\,\id\ (Fig.\,\ref{f-scarglehs0506}). The
strongest signal is found at $6.7706\pm0.0065$\,\id, which we identify
as the likely orbital period of HS\,0506+7725,
$\Porb\simeq212.7\pm0.2$\,min, where the error is determined from a
sine fit to the radial velocity data (Table\,\ref{t-rvfits}). The
Scargle periodogram computed from a fake data set results in a much
cleaner periodogram than that obtained from the observed data,
suggesting that the line wings are affected by additional velocity
contributions apart from the orbital motion.  The $\Ha$ radial
velocities folded over the orbital period (Fig.\,\ref{f-rvfolded})
display a low amplitude of \textbf{$42.6\pm4.4$\,\kms} and a
relatively large amount of scatter, again suggesting that the orbital
motion measured from the line wings is contaminated by another
velocity component.

\begin{table*} [t]
\caption[]{Sine fits to the \Ha\ emission line radial
  velocities\label{t-rvfits}. The methods employed were a convolution
  with a single Gaussian (SG) or \citeauthor{schneider+young80-2}'s
  (\citeyear{schneider+young80-2}) double-Gaussian prescription (DG).}
\begin{flushleft}
\begin{tabular}{ccccccc}
\hline\noalign{\smallskip} Object & Method & FWHM/Sep. (\kms) & \T\ & Period
(days) & K (\kms) & $\gamma$ (\kms)\\ \hline\noalign{\smallskip}
HS\,0139+0559 & SG & 600 & $2452998.985\pm0.013$ & $0.16923\pm0.00034$ &
$84.4\pm4.8$ & $18.4\pm3.5$ \\ 
HS\,0229+8016 & SG & 600 & $2452992.457\pm0.006$ & $0.161493\pm0.000034$ & 
$179.0\pm5.2$ & $35.5\pm3.5$ \\ 
HS\,0506+7725 & DG & 700/1500 & $2452990.679\pm0.022$ & $0.14770\pm0.00014$ & 
$42.6\pm4.4$ & $-66.7\pm3.0$\\
\noalign{\smallskip}\hline
\end{tabular}
\end{flushleft}
\end{table*}

\begin{figure}
\centerline{\includegraphics[angle=-90,width=8.8cm]{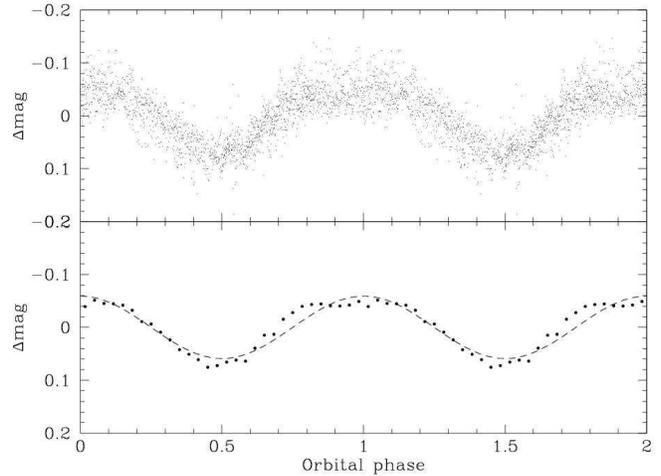}}
\caption[]{\label{f-lcfolded} HS\,0642+5049 photometric data from the
CAFOS (2004 October 9) and the IAC80 (2004 December 8 \& 9) folded on
$\Porb=225.90$\,min (top panel). The average light curve, binned
into 30 phase is shown in the bottom panel along with a sine fit
(dashed line).}
\end{figure}

Our time-series analysis of the photometry of HS\,0506+7725 did not
lead to the detection of any significant signal, either at long
(orbital) or at short (putative white dwarf spin) frequencies,
making the observed short-term variability (Fig.\,\ref{f-lc_hs0506}) a
nice example of non-coherent CV flickering. 

\subsection{HS\,0642+5049}
The 87 available spectra of HS\,0642+5049 were subjected to radial
velocity studies as outlined above, both using the single-Gaussian and
double-Gaussion convolution techniques. None of the resulting Scargle
periodograms contained any significant signal. Inspecting trailed
spectra assembled from our data, we concluded that HS\,0642+5049 does
not show any radial velocity variation at our spectral resolution. 

Considering the $\sim3.5$\,h modulation observed in the HS\,0642+5049
light curves, we used the the three longest and closest spaced photometric
data sets obtained at the Calar Alto 2.2\,m telescope (2004 October
25) at the IAC80 (2004 December 8 \& 9) to determine the orbital
period of the system. The strongest peak detected in the Scargle
periodogram computed from these data is found at
$6.3746\pm0.0065$\,\id, surrounded by one-day aliases
(Fig. \,\ref{f-scarglehs0642_phot}). In order to test the significance
of the signal, we created a set of fake data from a sine wave with a
frequency of 6.3746 evaluated at the exact times of the
observations. The alias structures of the periodograms calculated from
the fake data and the real data agree well. We conclude that the 
orbital period of HS\,0642+5049 is  $\Porb\simeq225.90\pm0.23$\,min.

Fig.\,\ref{f-lcfolded} shows the CAFOS and the IAC80 photometry folded
on 225.90 min and averaged into 30 phase bins which reflects the
morphology of the individual light curves (Fig. \,\ref{f-lc_hs0642}).

\section{Discussion}

\subsection{The inventory of the 3--4\,h orbital period range}
\begin{figure}
\centerline{\includegraphics[angle=-90,width=\columnwidth]{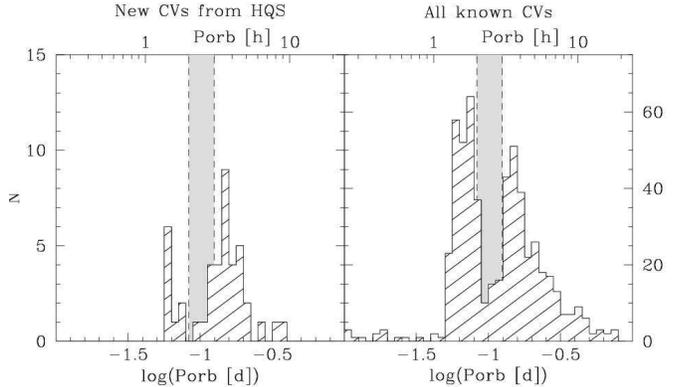}}
\caption[]{\label{f-porb} The period distribution of 42 new CVs
discovered in the HQS (left panel) and of all known CVs (right panel,
from \citealt{ritter+kolb03-1}, V7.4). The 2--3\,h period gap is shaded
in gray.}
\end{figure}

The primary aim of our search for CVs in the HQS is to establish the
orbital periods and CV subtypes for a large sample of CVs that were
selected in a homogenous way based on their spectroscopic
properties. The properties of this sample will then be compared with
the predictions of CV evolution theory. Here, we report the
spectroscopic identification and detailed follow-up studies of
HS\,0139+0559, HS\,0229+8016, HS\,0506+7725 and HS\,0642+5049, which
have the orbital periods of 243.69 min, 232.550 min, 212.7 min and
225.90 min, respectively. This follows the trend noticed by
\citet{gaensickeetal02-2} and more recently by \citet{gaensicke04-1}
and \citet{rodriguez-gil05-1} that the majority of the new CVs
identified in the HQS have orbital periods above the period gap and
the bulk of them are concentrated in the 3--4\,h orbital period
range. Currently, orbital periods have been determined for 42 systems
out of a total of 53 new HQS CVs, and Fig.\,\ref{f-porb} compares the
period distribution of these new HQS CVs with the period distribution of
the CVs from the \citeauthor{ritter+kolb03-1} catalogue
(\citeyear{ritter+kolb03-1}, V7.4). Even though the follow-up of the
new HQS CVs is not yet complete, it is already now clear that our
survey \textit{did not} identify the large number of short-period CVs
predicted by the population models \cite[e.g.][]{kolb93-1,
howelletal97-1}, even though our selection method (\,=\,detection of
Balmer emission lines) is most-suited for the identification of low
mass transfer systems, that might be inconspicuous in other ways
(variability, X-rays), such as e.g. the short-period dwarf novae
HS\,1449+6415 \citep{nogamietal00-1} and HS\,2219+1824
\citep{rodriguez-giletal05-1}, or the ultra-short period
HS\,2331+3905, which might be a WZ\,Sge type dwarf novae with
extremely long outburst recurrence times
\citep{araujo-betancoretal05-1}. The (somewhat preliminary) conclusion
is that if a large number of short-period CVs does indeed exist, they
must look different from the well-known examples such as e.g. WZ\,Sge.

\begin{figure}
\centerline{\includegraphics[angle=-90,width=\columnwidth]{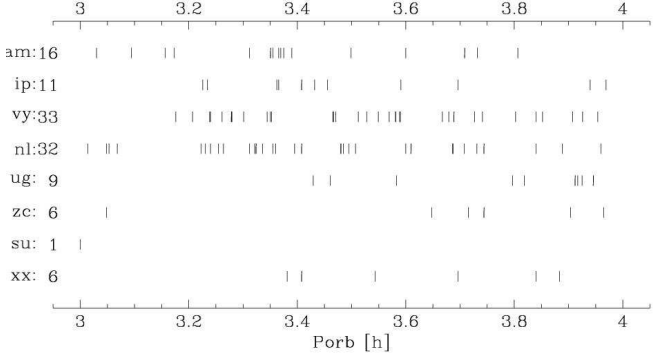}}
\caption[]{\label{f-cvtypes} The period distribution of the individual
CV subtypes in the 3--4\,h period range. From top to bottom: polars
(am), intermediate polars (ip), VY\,Scl and SW\,Sex stars (vy),
novalike variables and nova remnants that are neither VY\,Scl nor
SW\,Sex stars (nl), U\,Gem type dwarf novae (ug), Z\,Cam type dwarf
novae (zc), SU\,UMa type dwarf novae (su), and systems with
undetermined CV subtype (xx). 
}
\end{figure}

The HQS CV survey has been very prolific in identifying relatively
bright long-period CVs, with a distinct preference for the 3--4\,h
period range (Fig.\,\ref{f-porb}), including the four new CVs
presented in this paper. The majority of these new long period CVs are
weak or no X-ray emitters, and display little long-term
variability~--~in fact, only five confirmed dwarf novae are among the
28 new systems found above the gap. \citet{gaensicke04-1} and
\citet{rodriguez-gil05-1} pointed out the large frequency of
SW\,Sextantis stars among the new HQS CVs, which represent 25\% of all
newly identified CVs above the gap, and nearly half of all the new CVs
in the 3--4\,h period range. For comparison, we show in
Fig.\,\ref{f-cvtypes} the inventory of the 3--4\,h orbital period
range according to \citeauthor{ritter+kolb03-1}
(\citeyear{ritter+kolb03-1}, V7.4). We find that 114 CVs (20\% of all
CVs with known \Porb) inhabit the 3--4\,h period range, of which 27
(24\%) are confirmed magnetic systems (intermediate polars,
polars). 33 (29\%) belong to the group of either VY\,Scl or SW\,Sex
stars, which share similar properties, and are suspected to contain
magnetic white dwarfs as well \citep[e.g.][]{rodriguez-giletal01-1,
hameury+lasota02-1}. While the ratio of definite magnetic CVs in the
3--4\,h period range (24\%) is already very high compared to the
incidence of magnetism in isolated white dwarfs
\citep{liebertetal03-1}, a confirmation of significant magnetism in
the white dwarfs in VY\,Scl/SW\,Sex stars would raise the ratio of
magnetic/non-magnetic CVs well above 50\%, which is in conflict with
any of the current models of CV evolution. Whatever the verdict on the
magnetic fields in VY\,Scl/SW\,Sex stars will be, the large recurrence
of this CV subtype suggest that they represent an important phase of
CV evolution rather than some unusual combination in their physical
properties. For completeness, 32 (28\%) novalike variables that do not
belong to either the VY\,Scl or SW\,Sex class populate the 3--4\,h
period range\footnote{The classification of \citealt{ritter+kolb03-1}
is somewhat confusing, as their use of the UX\,UMa type disagrees with
the more common definition of systems charcterised by persistent broad
Balmer absorption lines.}. While a number of those systems definitely
do not share any of the VY\,Scl/SW\,Sex properties, a fair fraction of
these systems has been studied only in a very limited way, and hence
some of them may join the VY\,Scl/SW\,Sex class upon a more detailed
scrutiny. Finally, 16 (14\%) dwarf novae are known in the 3--4\,h
period range, the scarcity of systems undergoing thermal disc
instabilities just above the period gap is a well-known fact
\citep[e.g.][]{shafteretal86-1, shafter92-1}.


\begin{table*} [t]
\caption[]{\label{t-compsys} Comparison of the observational
  characteristics of the four new CVs. The (non)detection of X-ray
  emission referes to the ROSAT All Sky Survey
  \citep{vogesetal00-1}. The CV subtypes are abbreviated as
  UX\,=\,UX\,UMa type novalike variable, ZC\,=\,Z\,Cam type dwarf
  nova, VY\,=\,VY\,Scl star.}
\begin{flushleft}
\setlength{\tabcolsep}{1.3ex}
\begin{tabular}{cccccccc}
\hline\noalign{\smallskip} 
Object & \Porb & Radial velocity & 
\multicolumn{3}{c}{Photometric variability} & X-ray & type \\
& [min] & variation & Orbital & Flickering & Long-term \\
\hline\noalign{\smallskip}
HS\,0139+0559 & 243.7 & clean, moderate amplitude & none & none & none
& no & UX or ZC\\
HS\,0229+8016 & 232.6 & clean, large amplitude & very low amplitude &
none & $\sim1.3$\,mag & no & UX or ZC \\
HS\,0506+7725 & 212.7 & scatter, low amplitude & not obvious & large
amplitude & $\sim3$\,mag low state & yes & VY \\
HS\,0642+5049 & 225.9 & very low amplitude & moderate amplitude & low
amplitude & none &  maybe & UX or ZC \\
\noalign{\smallskip}\hline
\end{tabular}
\end{flushleft}
\end{table*}

\subsection{The nature of the four new CVs}
Based on observational characteristics (summarised in
Table\,\ref{t-compsys}) we discuss the likely nature of the four new
CVs.

HS\,0506+7725 shows short time scale flickering with quasi-periodic
oscillations on time scales of $\sim$15 min. The relatively narrow
emission lines and the low amplitude of the radial velocity variations
suggest a low inclination. The system has been detected in the RASS
\citep{vogesetal00-1} at $0.07\,\mathrm{cts\,s^{-1}}$
(\object{1RXS\,J051336.1+772836}) with a hard spectrum, and has been
previously detected as an X-ray source by EINSTEIN
(\object{2E\,0506.1+7725}). The presence of moderately strong
\Line{He}{II}{4686} emission in the identification spectrum
independently confirms the presence of ionising radiation in the
system. The detection of a deep low state at $B\simeq18.3$ on one of
the HQS prism plates clearly identifies the system as a VY\,Scl
star. The system does not display \textit{at face value} evidence for
being an SW\,Sex star, but being obviously a low-inclination binary, a
spectroscopic study at higher resolution would be useful to test for
anomalous radial velocity behaviour in the emission lines.

The other three systems, HS\,0139+0559, HS\,0229+8016, and
HS\,0642+5049 are spectroscopically very similar, being characterised
by thick-disc absorption line spectra. The fact that we have observed
them on various occasions, and found them always at nearly the same
magnitude and with the same spectral properties\footnote{With one
exception: HS\,0229+8016 has been observed in August 1992 in a
somewhat fainter state, $V\simeq15.0$, compared to its typical
brightness near 14\,mag. During that occasion, the Balmer and
\Ion{He}{I} absorption lines were absent/weak, and the strength of the
emission lines had markedly increased.} makes it very unlikely that
these systems are U\,Gem-type dwarf novae observed during
outburst. While HS\,0139+0559 and HS\,0229+8016 are not detected in
the RASS, a faint X-ray source is found near HS\,0642+5049
(\object{1RXS\,J066618.4+504601}, $0.02\,\mathrm{cts\,s^{-1}}$) which
coincides within the 29\arcsec position error of the RASS detection
with the CV. The fact that there are no other nearby objects suggests
that HS\,0642+5049 is a weak X-ray emitter. None of the systems
shows strong flickering activity. One puzzling difference among the
three systems is that whereas HS\,0139+0559 and HS\,0229+8016 show no or only
very low-amplitude orbital photometric variability, but exhibit clean
quasi-sinusoidal radial velocity variations in their emission lines,
HS\,0642+5049 does not display any radial velocity variation, but a
0.2\,mag photometric modulation. It is very difficult to reconcile
this opposite difference in spectroscopic/photometric behaviour in the
simple picture of a high-mass transfer CV with a steady-state
accretion disc. Based on our data, we identify all three systems
either as UX\,UMa-type novalike variables, or as Z\,Cam-type dwarf
novae observed in periods of standstill. Optical long-term monitoring
will be necessary to distinguish between these two possibilities.

\section{Conclusions}
We have identified HS\,0139+0559, HS\,0229+8016, HS\,0506+7725 and
HS\,0642+5049 as long-period CVs with the orbital periods of
$243.69\pm0.49$ min, $232.550\pm0.049$ min, $212.7\pm0.2$ min and
$225.90\pm0.23$, respectively. HS\,0506+7725 is a VY\,Scl novalike
variable characterised by a strong emission-line spectrum.
HS\,0139+0559, HS\,0229+8016 and HS\,0642+5049 have thick-disc spectra
and are either UX\,UMa type novalike variables or Z\,Cam dwarf
nova. None of the objects is a strong X-ray source or displays
large-amplitude outbursts, which underlines the strength of  CV
surveys on spectroscopically selected candidates. 

\acknowledgements AA thanks the Royal Thai Government for a
studentship. BTG and PRG were supported by a PPARC Advanced Fellowship
and a PDRA grant, respectively. The HQS was supported by the Deutsche
Forschungsgemeinschaft through grants Re\,353/11 and Re\,353/22. Tom
Marsh is acknowledged for developing and sharing his reduction and
analysis package \texttt{MOLLY}. We thank the referee, Mike Shara, for
his comments that lead to an improved presentation of the paper.

\bibliographystyle{aa}
\bibliography{aamnem99,aabib}

\end{document}